\documentclass[twocolumn,bibnotes 
]{revtex4-1}
\usepackage{amssymb}
\usepackage{graphicx}
\usepackage{amsmath,rotating,amscd,bm}
\usepackage{color,ulem}

\def\p{\partial}

\def\underline{\underline}

\def\=:{=\hspace{-.7em}\raisebox{1.1ex}{.}\hspace{.1em}\raisebox{-0.2ex}{.} }

\newcommand {\be}{\begin{eqnarray}}
\newcommand {\ee}{\end{eqnarray}}

\begin{document}

\begin{flushright}YGHP-21-05
\end{flushright}

\title{Phases of rotating baryonic matter: 
non-Abelian chiral soliton lattices,
\\
antiferro-isospin chains, 
and ferri/ferromagnetic magnetization
}

\author{Minoru Eto$^{1,2}$, Kentaro Nishimura$^3$, Muneto Nitta$^{2,4}$}

\affiliation{%
$^1$Department of Physics, Yamagata University, 
Kojirakawa-machi 1-4-12, Yamagata,
Yamagata 990-8560, Japan, \\
$^2$Research and Education Center for Natural Sciences, Keio University, 4-1-1 Hiyoshi, Yokohama, Kanagawa 223-8521, Japan\\
$^3$Department of Physics, Keio University, 3-14-1 Hiyoshi, Yokohama, Kanagawa 223-8522, Japan, \\
$^4$Department of Physics, Keio University, 4-1-1 Hiyoshi, Yokohama, Kanagawa 223-8521, Japan
}%

\date{\today}

\begin{abstract}

A chiral soliton lattice (CSL), proposed 
as the ground state of rotating baryonic matter at a finite density, is shown to be unstable in a large parameter region for two flavors owing to pion condensations, leading to two types of non-Abelian (NA) CSL phases (dimer and deconfining phases). We determine the phase diagram where the dimer phase meets the other phases and QCD vacuum at three tricritical points. The critical angular velocity for NA-CSLs is lower than the $\eta$-CSL. Each NA soliton carries an isospin, and an antiferro-isospin chain is formed leading to gapless isospinons. The anomalous coupling to the magnetic field provides the NA-CSL ($\eta$-CSL) with a ferrimagnetic (ferromagnetic) magnetization.

\end{abstract}

\maketitle

{\it Introduction.} 
Determination of the quantum chromodynamics (QCD) phase diagram under extreme conditions, such as at finite temperature and/or density, is a crucial problem in elementary particle physics, nuclear physics, and astrophysics. 
It has been reported that quark-gluon plasmas produced in non-central heavy-ion collision experiments at the Relativistic Heavy Ion Collider (RHIC) have the largest vorticity observed thus far, of the order of $10^{22}/{\rm s}$ \cite{STAR:2017ckg,STAR:2018gyt}. Thus, rotating QCD matter has received significant attention in recent years.
Moreover, recent developments in neutron star observations, 
such as the Laser Interferometer Gravitational-wave Observatory (LIGO) merger for the observation of gravitational waves from a neutron star
\cite{TheLIGOScientific:2017qsa,Abbott:2020uma} 
and the Neutron star Interior Composition Explorer (NICER) mission~\cite{Riley:2019yda,
  Miller:2019cac}  
may reveal states of QCD matter realized in rapidly rotating neutron stars. 
Therefore, it is crucial to investigate the effects of rotation on QCD matter. These effects have been theoretically studied by several researchers
\cite{Chen:2015hfc,Ebihara:2016fwa,Jiang:2016wvv,Chernodub:2016kxh,Chernodub:2017ref,Liu:2017zhl,Zhang:2018ome,Wang:2018zrn,Chen:2019tcp,Chen:2021aiq,Huang:2017pqe,Nishimura:2020odq};
in particular, 
it has been predicted 
that due to the chiral vortical effect (CVE) \cite{Vilenkin:1979ui,Vilenkin:1980zv,Son:2009tf,Landsteiner:2011cp,Landsteiner:2012kd,Landsteiner:2016led},
baryonic matter under rapid rotation 
exhibits a chiral soliton lattice (CSL),
which is a periodic array of topological solitons that spontaneously break 
a translational symmetry 
\cite{Huang:2017pqe,Nishimura:2020odq}.	
Similar CSLs also appear in QCD under
an external magnetic field \cite{Son:2007ny,Eto:2012qd,Brauner:2016pko,Chen:2021vou}
and thermal fluctuation \cite{Brauner:2017uiu,Brauner:2017mui,Brauner:2021sci}
(see also \cite{Yamada:2021jhy,Brauner:2019aid,Brauner:2019rjg}).
More generally, 
CSLs universally appear in various condensed matter systems; a partially twisted coherent spin structure in chiral magnets is realized as a CSL \cite{Dzyaloshinsky:1964dz,togawa2012chiral}.
Notably, as an important nanotechnological application, 
information processing using CSL has the potential to improve the performance of magnetic memory storage devices and magnetic sensors \cite{togawa2016symmetry}.

In this Letter, the phase diagram of rotating QCD matter is determined
with two-flavor quarks at finite baryon chemical potential,  
indicating that 
the ground state of the QCD at finite density under sufficiently fast rotation is a novel inhomogeneous state, called a {\it non-Abelian (NA) CSL}.
It is noted that the conventional CSL of $\eta$ meson 
($\eta'$ for three flavors)
\cite{Huang:2017pqe,Nishimura:2020odq},
referred to as the $\eta$-CSL, 
is unstable against the pion's fluctuations, 
leading to spatially modulated pion condensations 
in a large parameter region.
There, in addition to 
the usual phonon,
the vector $SU(2)_{\rm V}$ symmetry
is also spontaneously broken 
around each constituent soliton, 
resulting in the localization of $S^2$
NA Nambu-Goldstone (NG) modes (an isospin), 
thus named an NA chiral soliton (CS) 
\cite{Nitta:2014rxa,Eto:2015uqa}.  
The NA-CSL phase can be further divided into two phases: 
the dimer phase, in which the two solitons form a dimer molecule 
of a soliton pair with opposite isospins, 
and the deconfining phase, in which 
they repel each other completely, thus forming 
an equally-separated opposite-isospin soliton lattice. 
Furthermore, gapless NG modes ``isospinons'' propagate 
along the lattice direction, 
analogous to magnons in antiferromagnets.
The CSL also shows magnetization 
due to the anomalous coupling of the $\pi_0$ meson to the magnetic field \cite{Son:2004tq,Son:2007ny,Eto:2013hoa}; 
the NA-CSL ($\eta$-CSL) is ferrimagnetic (ferromagnetic).
We  discuss a possibility to find the CSL in 
future low-energy heavy ion collision experiments.

{\it Model.} 
The two-flavor chiral Lagrangian of the $\eta$ meson and pions $\vec \pi$ 
is given by
\be
{\cal L} &=& 
\frac{\Omega\mu_{\rm B}^2}{2\pi^2N_c}\p_z\frac{\eta}{f_\eta} +
\frac{f_\pi^2}{4} g^{\mu\nu} \, {\rm tr} \left[\p_\mu \Sigma \p_\nu \Sigma\right] 
+ \frac{1}{2}g^{\mu\nu}\partial_\mu\eta\partial_\nu \eta \nonumber\\
&+& \left\{\frac{A}{2}\left(\det U -1\right)
+ \frac{B}{2}\, {\rm Tr}\left[M(U -1)\right] + {\rm h.c.}\right\},
\ee
where the $U(2)$ field $U$ is decomposed as 
$U = \Sigma \exp(i\eta/f_{\eta})$ with the $SU(2)$ element
$\Sigma = \exp(i\tau_A\pi_A/f_\pi)$ ($\tau_A$ is Pauli's matrix $A = 1, 2, 3$), and 
$f_{\pi,\eta}$ are the decay constants of the pions and $\eta$ meson.
This system is put under rotation about the $z$-axis with angular velocity $\Omega$.
The rotation effect is considered in the metric tensor $g_{\mu\nu}$, 
and the first term $\mu_{\rm B}$ (baryon chemical potential) 
reproduces the CVE in terms of the $\eta$ meson.\footnote{See 
Appendix \ref{app:a}
for details on the CVE term.
} The fourth and fifth terms reflect the QCD anomaly and quark mass term, respectively.
It is assumed that $M = m {\bf 1}_2$, which is valid for $\mu_{\rm B} \gg m_{\rm u,d}$.
The Lagrangian is invariant under the vector symmetry $SU(2)_{\rm V}$:
$U \to V U V^\dagger$, while the $U(1)_{\rm A}$ (chiral) symmetry is explicitly broken by the anomaly (mass) term.

This study considers an array of solitons, namely CSLs,
extending along the rotation axis $z$.
For this purpose, it is assumed that the fields depend on
the $z$-coordinate alone, without loss of generality, 
and $\pi_{1, 2} = 0$ is set for constructing ground states. 
Furthermore, dimensionless fields $\phi_{0,3}$ defined by
$\phi_0 = \eta/f_{\eta}$ and 
$\phi_3 = \pi_3/f_\pi$
are largely used, along with the dimensionless variables
$\zeta = \sqrt{C} z/ f_\eta$,
$\epsilon \equiv 1 - ( f_\pi/f_{\eta})^2$, and
$S = \Omega/(2\pi^2N_cf_\eta\mu_{\rm B}^{-2}\sqrt C)$
with $C = \sqrt{A^2 + (4mB)^2}$.
Then, the reduced Hamiltonian density is
$
\frac{{\cal H}}{C} = \frac{1-\epsilon}{2}\phi_3'{}^2
+ \frac{1}{2}\phi_0'{}^2
+ \sin \beta(1-\cos2\phi_0) 
+ \cos\beta\left(1-
\cos\phi_0\cos\phi_3
\right) 
- S\phi_0'$,
where $A$ and $B$ are parameterized by $\tan\beta = A/(4mB)$.
It can be observed that the independent parameters are $\epsilon$,
$\beta$, and $S$.
The potential minimum is unique as $U = {\bf 1}_2$, which corresponds to
$(\phi_0, \phi_3) = (2m\pi, 2n\pi)$ and $((2m + 1)\pi,(2n + 1)\pi)$,
with $m, n \in \mathbb{Z}$ in the $\phi_0$-$\phi_3$ plane.\footnote{
See 
Appendix \ref{app:a} 
for the potential and vacuum with various parameter choices.
}

{\it Linear instability of $\eta$-CSL.} 
Recently, Ref.~\cite{Nishimura:2020odq} reported that when the rotation 
speed $S$ is sufficiently large, $\eta$-CSL, which straightly
connects the infinite discrete vacua $(\phi_0,\phi_3) = (2\pi n, 0)$
(where $\phi_0$ is a monotonically increasing step-function-like configuration, whereas $\phi_3$
vanishes everywhere), becomes the ground state owing to the CVE term.
\begin{figure}[t]
\begin{center}
\includegraphics[width=0.48\textwidth]{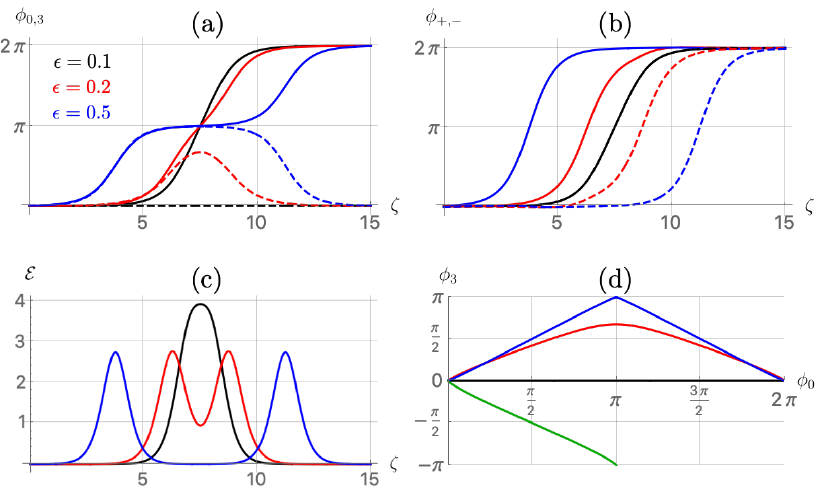}
\caption{Numerical solutions of CSLs 
(shown within one period) for $\beta = \pi/16$ and $\ell = 15$.
The black, red, and blue curves correspond to $\epsilon = 0.1$ ($\eta$-CSL), $0.2$ (dimer NA-CSL),
and $0.5$ (deconfined NA-CSL), respectively. The solid (dashed) curves show (a) $\phi_0$ ($\phi_3$) and (b) $\phi_+$ ($\phi_-$). (c) Energy densities without the CVE term.
(d) Solutions plotted in the $\phi_0$-$\phi_3$ plane. The green curve indicates the single d-CS with $\epsilon = 0.2$.
}
\label{fig:configurations}
\end{center}
\end{figure}

First, the linear (in)stability of the $\eta$-CSL, which was not studied in Ref.~\cite{Nishimura:2020odq}, is clarified. 
Let $\bar\phi_0(\zeta)$ be the $\eta$-CSL background solution with lattice size $\ell$, where $\bar\phi_A(\zeta) = 0$ ($A = 1, 2,3$).
The black curves in Fig.~\ref{fig:configurations} represent a typical configuration (with a single period) for $(\beta, \epsilon, \ell) = (\pi/16, 0.1, 15)$.
All fields are perturbed as $\phi_0 = \bar\phi_0 + \delta \phi_0$
and $\phi_A = \delta \phi_A$ 
and are not mixed in the linearized equations of motion (EOMs) \footnote{See 
Appendix \ref{app:b}
for the derivations of the linearized EOMs for the $\eta$-CSL.}. 
Clearly, no instability arises 
in the $\delta \phi_0$ sector,
and a tachyonic instability  may exist in the $\delta\phi_A$ sector.
It is found that $\eta$-CSL is always stable for $\epsilon < 0$ 
($f_\eta < f_\pi$), 
whereas it can be unstable for $\epsilon > 0$ ($f_\eta > f_\pi$).\footnote{
In the three-flavor case \cite{Nishimura:2020odq}, the ratio $f_{\eta'}/f_\pi$ for the vacuum values is estimated
as $\approx 1.1$, implying that $\epsilon \approx 0.17 > 0$.
}
\begin{figure}[t]
\begin{center}
\includegraphics[width=0.48\textwidth]{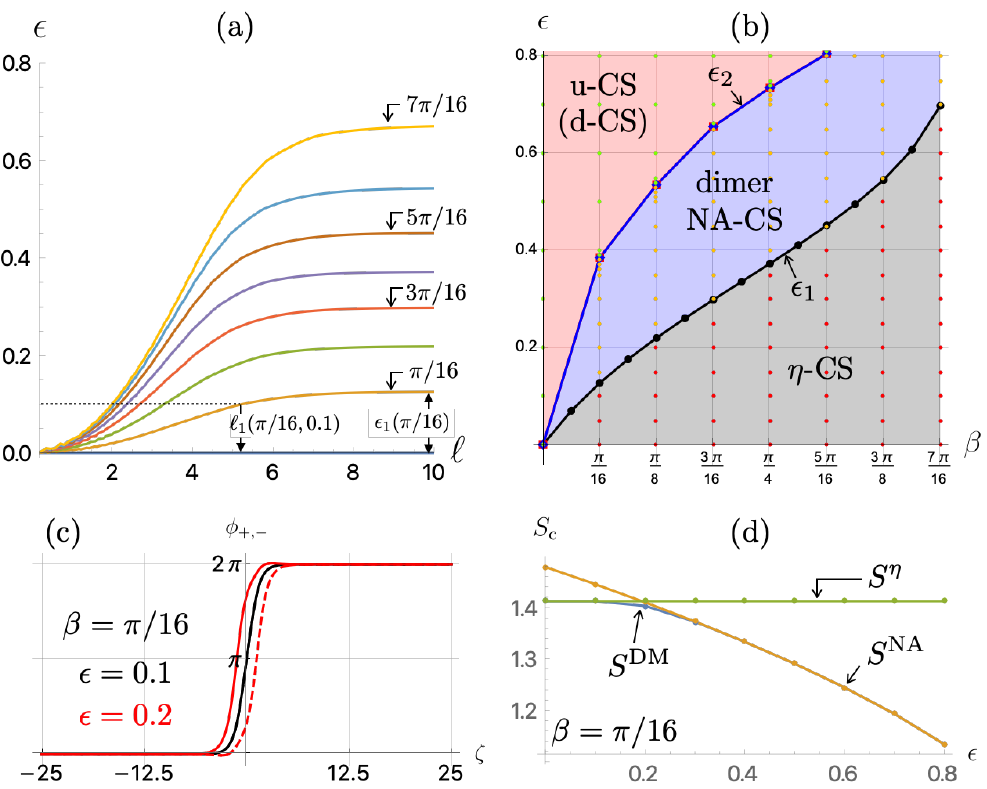}
\caption{
(a) Linear stability of the $\eta$-CSL in the $\ell$-$\epsilon$ plane. 
It is unstable above each curve with $\beta = 0$ -- $7\pi/16$.
(b) Phase diagram of a single CS.
The black and blue dotted lines show $\epsilon_1$ and $\epsilon_2$, respectively.
(c) The solid (dashed) curve corresponds to $\phi_+$ ($\phi_-$). The black (red) curves are as follows: 
For $\eta$-CS of $\epsilon = 0.1$ (dimer NA-CS of $\epsilon = 0.2$). $\beta = \pi/16$.
(d) Three critical velocities $S^\eta $, $S^{\rm NA} $, and $S^{\rm DM}$
are shown for $\beta = \pi/16$.
}
\label{fig:localstability}
\end{center}
\end{figure}
Figure~\ref{fig:localstability}(a) 
shows stable/unstable regions of $\eta$-CSL
in the parameter plane $\ell$-$\epsilon$.
Each curve shows a boundary above (below) at which the lowest energy eigenvalue
is negative (positive), leading to its instability (stability) 
for fixed $\beta$ values.
The curves starting from $(\ell,\epsilon) = (0,0)$ 
monotonically increase for small $\ell$ 
and tend to critical value $\epsilon_1$ 
when the period $\ell$ exceeds a typical size of the single soliton, which is
of order one with respect to the dimensionless coordinate $\zeta$.
The $\eta$-CSL is reduced to a single $\eta$-CS for a large period $\ell = \infty$.
The value $\epsilon_1$ shown in Fig.~\ref{fig:localstability}(b)
is defined as the border of the linear stability of a single $\eta$-CS.
For a given pair of $\beta$ and $\epsilon$, if $\epsilon > \epsilon_1$, 
there exist no stable $\eta$-CSLs. If $\epsilon < \epsilon_1$,
$\eta$-CSL is stable only when $\ell$ is larger than the critical value
$\ell_1 = \ell(\beta,\epsilon)$ on the boundary, however, it is unstable
below $\ell_1$, see Fig.~\ref{fig:localstability}(a) for  $\ell_1(\pi/16,0.1)$
as an example.
Namely, a sparse (dense) $\eta$-CSL tends to be (un)stable.

{\it Single-CS and critical angular velocity.} 
Here, by examining a single CS, the critical angular velocity $S_{\rm c}$ (dimensionless) is determined, above which the CSL becomes the ground state.

First, consider a single $\eta$-CS ($\phi_3 = 0$) that connects $(\phi_0,\phi_3) = (0,0)$ and $(2\pi,0)$,
as $\zeta = -\infty$ and $+\infty$ (see the black segment in Fig.~\ref{fig:configurations}(d)).
Because the endpoints are identical ($U = {\bf 1}_2$), this is a loop in the field space.
The loop is topologically nontrivial because it winds once around $U(1) \subset U(2)$.
The solution is easily obtained because $\phi_3 = 0$ can be consistently set, following which
EOM reduces to the well-known double sine-Gordon equation. In Ref.~\cite{Nishimura:2020odq}, it was found that the tension  
of $\eta$-CS (integration of ${\cal H}$ without the CVE term for $-\infty \le \zeta \le \infty$) is given by
$T^\eta (\beta) = 4\int^\pi_0\sqrt{\sin\beta\sin^2\theta 
+ \cos\beta\sin^2\frac{\theta}{2}}\,d\theta$, and the total tension including the CVE term is $M^\eta = T^\eta - 2\pi S$. 
Because of the CVE term, $M^\eta $ reduces as $S$ increases.
When $S$ is equal to the critical value $S^\eta  = T^\eta /2\pi$,
the single $\eta$-CS is degenerate in energy with the homogeneous QCD vacuum.
Note that the $U(2)$ field $U$ is proportional to ${\bf 1}_2$; therefore,
$SU(2)_{\rm V}$ is unbroken everywhere. 

There exists another loop 
topologically distinguishable from a single $\eta$-CS. 
It connects $(0,0)$ and $(\pi,\pi)$ at $\zeta = -\infty$ and $+\infty$; thus, it corresponds to
a diagonal curve between $(0,0)$ and $(\pi,\pi)$ in Fig.~\ref{fig:configurations}(d).
This winds only a half around $U(1)$. In this sense, it is
the minimum loop 
among topologically nontrivial loops.
This is possible owing to the $\mathbb{Z}_2$ quotient of 
$U(2) \simeq [U(1)\times SU(2)]/\mathbb{Z}_2$; specifically, $e^{i\phi_0} = -1$ and 
$\Sigma = e^{i\pi\tau_3} = -{\bf 1}_2$ give $U = {\bf 1}_2$ at $(\pi,\pi)$.
As $\phi_3$ is no longer zero, the corresponding soliton is 
a NA-CS \cite{Nitta:2014rxa,Eto:2015uqa}.
Indeed, an important feature specific to NA-CS is that 
it possesses NA moduli; it passes the point $(\pi/2,\pi/2)$,
namely $U = -\tau_3$, where $SU(2)_{\rm V}$
is spontaneously broken into its subgroup $U(1)$.
Therefore, the NG modes $SU(2)_{\rm V}/U(1) \simeq S^2$ appear locally around the NA-CS, endowing it with an isospin.

Consider this NA-CS as the north pole of the moduli space $S^2$. 
By applying the $SU(2)_{\rm V}$ transformation 
to the north pole solution, 
a continuous family of solutions having the same tension is obtained. 
In the $\phi_0$-$\phi_3$ plane, the NA-CS connecting $(0,0)$ and $(\pi,-\pi)$, (corresponding to the south pole) can be observed (see the green curve in Fig.~\ref{fig:configurations}(d)). 
The former is referred to as u-CS and the latter as d-CS.
All other solutions connect two points outside this plane.
Let $T^{\rm NA} $ be the tension of the single NA-CS (obtained from ${\cal H}$ with the CVE term excluded). 
It is dependent on both $\beta$ and $\epsilon$ and can only be computed numerically.
The total tension, including the CVE term, is $M^{\rm NA} = T^{\rm NA} - \pi S$.
Note that the second term is $\pi S$ and not $2\pi S$ because $\phi_0$ increases by $\pi$ for NA-CS. Hence, the other critical angular velocity is obtained as $S^{\rm NA} =T^{\rm NA} /\pi$. 

The final possibility is a dimer state, {i.~e.~}
a pair of the u-CS connecting $(0,0)$ and $(\pi,\pi)$ 
and d-CS connecting $(\pi,\pi)$ and $(2\pi,0)$.
This pair is topologically equivalent
to a single $\eta$-CS.
Whether they split or combine 
is determined by 
their interactions, depending on 
parameters $\epsilon$ and $\beta$. 
Qualitatively speaking, it is found that a positive (negative) $\epsilon$ value induces a repulsion (attraction) between them, 
while a non-zero $\beta$ value yields an attractive interaction.
When the repulsive and attractive
interactions balance, u- and d-CSs are bounded to form a dimer with
a finite distance.
Let $T^{\rm DM} $ be the tension of the dimer obtained from ${\cal H}$ without the CVE term.
Then, the critical angular velocity is given by $S^{\rm DM} = T^{\rm DM} /2\pi$.
However, when the attractive force dominates, u- and d-CSs coalesce, reducing to a single $\eta$-CS.
When the repulsive force dominates, they repel each other, and the most stable state is (infinitely separated) single NA-CSs.

Thus, three critical velocities were found: $S^\eta $, $S^{\rm NA} $, and $S^{\rm DM}$.
The actual critical angular velocity is given by 
$S_{\rm c}(\beta,\epsilon) \equiv \min\left[S^{\eta}, S^{\rm NA}, S^{\rm DM}\right]$.
The CS that is realized
depends on $\beta$ and $\epsilon$.

The existence of all three cases was confirmed by numerically solving EOMs.\footnote{See 
Appendix \ref{app:a}
for the derivation of the EOMs.}
A relaxation method  was applied with an initial configuration of a pair of separated u- and d-CSs.
Whether the solution is Abelian can be easily determined by plotting $\phi_{\pm} = \phi_0\pm\phi_3$ because a $2\pi$-jumping soliton of $\phi_+$ ($\phi_-$) represents u-(d-)CS.
Hence, when a convergent
configuration has $\phi_+$ and $\phi_-$ which lie on top of each other, it can be concluded that the ground state is $\eta$-CS; otherwise, it is NA-CS. Figure ~\ref{fig:localstability}(c) shows examples of $\eta$-CS and NA-CS.
Then, a dimer and a pair of repelling u- and d-CSs were further distinguished by observing their separation.
The result was superposed on Fig.~\ref{fig:localstability}(b).
The red dots, which represent $\eta$-CS, are all below $\epsilon_1$,
indicating a consistent relationship between the linear stability and relaxation analysis. 
The yellow dots correspond to dimers, and the green dots represent a repelling pair of u- and d-CSs. 
$\epsilon_1$, which was originally introduced as the border of the linear stability of a single $\eta$-CS, is now identified with the boundary between a single $\eta$-CS and the dimer of u- and d-CSs.
Furthermore, another critical value is found,
$\epsilon_2$ (yellow curve with squares), which is the boundary between
a dimer and a single NA-CS.

Figure~\ref{fig:localstability}(d) shows 
the critical angular velocities for $\beta = \pi/16$
and $\epsilon \ge 0$.
It was found that $S_{\rm c} = S^\eta $ for $\epsilon < \epsilon_1$,
$S_{\rm c} = S^{\rm DM} $ for $\epsilon_1 < \epsilon < \epsilon_2$,
and $S_{\rm c} = S^{\rm NA} $ for $\epsilon > \epsilon_2$.
It should be noted that when NA-CS is the ground state, irrespective of whether it is a dimer or single NA-CS, $S_{\rm c}$ is lower than that of the $\eta$-CS found in Ref.~\cite{Nishimura:2020odq}. 
The $\eta$-CS and dimer are topologically indistinguishable. 
However, $SU(2)_{\rm V}$ is unbroken for the $\eta$-CS 
and is spontaneously broken into $U(1)_{\rm V}$ for the NA-CS.
The number of NG modes is $1$ (translation) for the $\eta$-CS 
and $1+2$ (translation and isospin) for the NA-CS.

\begin{figure}[t]
\begin{center}
\includegraphics[width=0.48\textwidth]{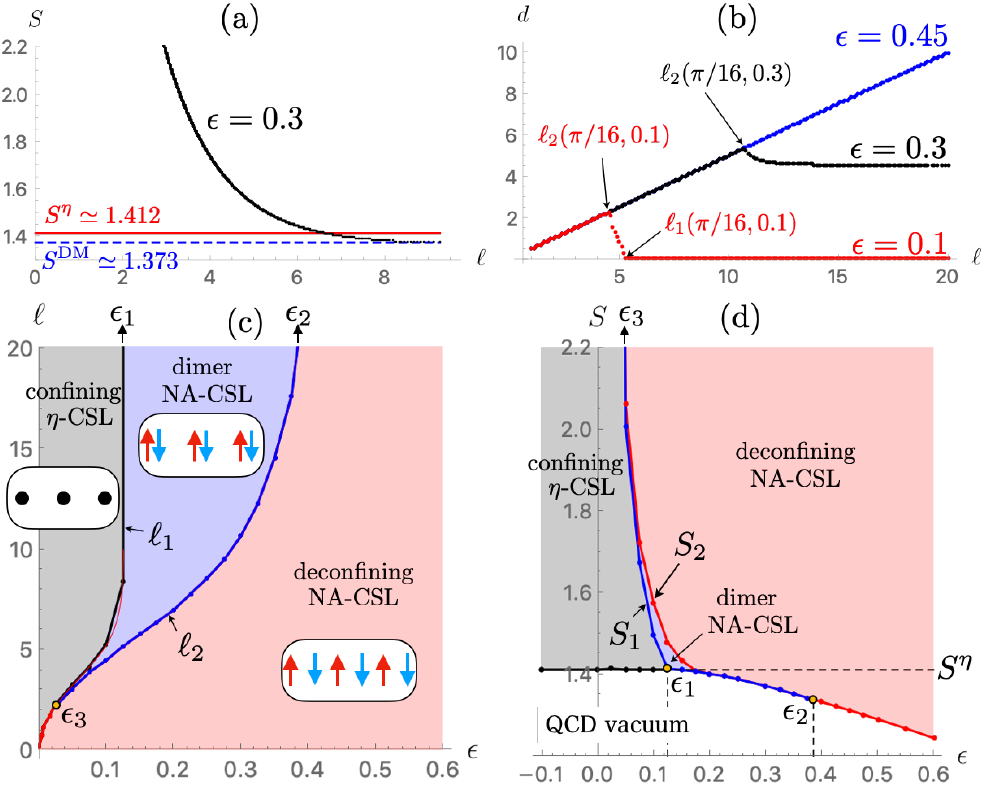}
\caption{CSLs for $\beta = \pi/16$.
(a) Relation $\ell(S)$ is shown for $\epsilon = 0.3$. 
$\ell$ diverges at $S = S^{\rm DM}$ $(< S^\eta)$.
(b) Distance $d$ between u- and d-CSs is shown as a function of $\ell$
for $\epsilon = 0.1$, $0.3$, and $0.45$. Points on the line $d = \ell/2$ correspond to the deconfined NA-CSL, the dimer NA-CSL ($d\neq0$), or the $\eta$-CSL ($d=0$).
(c) CSL phase diagram in the $\epsilon$-$\ell$ plane.
The array of arrows represents the antiferro-isospin chain of the NA-CSL, while that of black dots
represents the $\eta$-CSL (isoscalar).
The thin-red-solid curve corresponds to the border of 
the $\eta$-CSL linear stability 
for $\beta = \pi/16$, given in Fig.~\ref{fig:localstability}(a). 
(d) CSL phase diagram in the $\epsilon$-$S$ plane.
In (c) and (d), 
$\epsilon_{1, 2, 3}$ correspond to the three tricritical points.
}
\label{fig:CSL}
\end{center}
\end{figure}

{\it CSL for $S \ge S_{\rm c}$.} 
When $S$ exceeds $S_{\rm c}$, a periodic array of CSs appears with lattice spacing $\ell$.
Again, EOMs were numerically solved without
assuming $\phi_3 = 0$. However, in this case, periodic boundary conditions
$\phi_0(\zeta) = \phi_0(\zeta+\ell)+2\pi$
and $\phi_3'(\zeta)=\phi_3'(\zeta+\ell)$ are imposed.
Thus, $\ell$ is included as a free parameter in addition to $\beta,\epsilon$, and $S$, and determined as follows:
As $S$ appears only through the topological term, 
it does not appear in EOMs. Hence, the EOMs are first solved for various values of $\ell$ by setting specific values for $\beta$ and $\epsilon$.
Then the tension
$M(\ell,S;\epsilon,\beta) = \int^\ell_0 C^{-1}{\cal H}\,d\zeta = T(\ell;\epsilon,\beta) - 2\pi S$ is calculated,
where $T$ is the integration of the right-hand side of ${\cal H}$,
with the exception of the last term. Finally, $S$ is set, $M$ is regarded as a function of $\ell$, and 
a value of $\ell$ that minimizes the averaged mass
$\bar M(\ell) \equiv M(\ell;\epsilon,\beta,S)/\ell$ is considered.
Thus, $\ell(S;\beta,\epsilon)$ is obtained as a function of $S$ 
for specific $\beta$ and $\epsilon$ values. 
See Fig.~\ref{fig:CSL}(a) for an example of $(\beta,\epsilon) = (\pi/16, 0.3)$.

By repeating the above procedure for various $(\epsilon,\beta)$, the CSL phases could be clarified.
Similar to the single CSs explained above, CSLs are classified as Abelian or non-Abelian.
In a single CSL period, there exists a pair of u- and d-CSs.
In $\eta$-CSLs, u- and d-CSs are confined, whereas they are split for NA-CSLs.
The latter is further classified based on whether u- and d-CSs are bound to form a dimer.
Figure~\ref{fig:configurations}
shows examples of $\eta$-CSL (black for $\epsilon = 0.1$), dimer NA-CSL (red for $\epsilon = 0.2$),
and deconfined NA-CSL (blue for $\epsilon = 0.5$) for $\beta = \pi/16$ and $\ell = 15$.\footnote{See 
Appendix \ref{app:a}
for the numerical solutions with various parameters.}
The following three phases can be defined: 
(i) Confining phase: u- and d-CSLs are confined to form an $\eta$-CSL.
(ii) Dimer phase: u- and d-CSLs are confined and locally split to form a dimer.
(iii) Deconfining phase: They repel each other completely, thus forming 
an equally-separated up-and-down soliton lattice.  
The phase that is realized depends not only on $\beta$ and $\epsilon$, but also on $\ell$ (or $S$ from the relation $\ell(S)$).
Figure~\ref{fig:CSL}(b) shows a distance $d$ between u- and d-CSs in one period for $\beta = \pi/16$ and $\epsilon = 0.1$, $0.3$, and $0.45$.
As mentioned previously, the interaction between u- and d-CSs originating from $\epsilon (> 0)$
is repulsive, and that from $\beta$ is attractive.
This explains the behavior of $d$ at an asymptotically large period $\ell$.
The CSL at a large period $\ell$ is an $\eta$-CSL ($d = 0$) 
for $\epsilon = 0.1$ because the attractive force is dominant.
In addition, it is a dimer NA-CSL for $\epsilon = 0.3$ because $d$ tends to be constant, implying a dimer size for a large period $\ell$. 
The separation $d$ at $\epsilon = 0.45$ is on the line $d = \ell/2$, 
implying the u- and d-CSs
are maximally (and thus equally) separated in one period $\ell$,
for which the CSL belongs to the deconfining phase.
Note that these asymptotic behaviors are consistent with those depicted in Fig.~\ref{fig:localstability}(b),
which represents a single $\eta$-CS/NA-CS $(\ell\to\infty)$.
As $\ell$ decreases, a dimer NA-CSL ($\epsilon = 0.3$) at an asymptotically large $\ell$ 
enters the deconfining phase. The transition point $\ell_2$ can be understood as a point below
which mutual influence between adjacent dimers becomes significant.
Similarly, an $\eta$-CSL at an asymptotically large $\ell$ exhibits two successive transitions for a smaller $\ell$. 
The first transition from the $\eta$-CSL to the dimer NA-CSL 
 occurs at $\ell_1$, 
which is in good agreement with the critical period separating 
stable/unstable $\eta$-CSLs obtained via linear stability analysis (see $\ell_1$ values 
in Fig.~\ref{fig:localstability}(a) and \ref{fig:CSL}(b). 
Subsequently, the second transition occurs at $\ell = \ell_2$ ($< \ell_1$) from the dimer to
the deconfined NA-CSL. 

Figure~\ref{fig:CSL}(c) shows the phase diagram in the $\epsilon$-$\ell$ plane for $\epsilon > 0$ 
($\epsilon < 0$ is uniformly within the confining phase).
The curve $\ell_1$ divides the confining and NA-CSL phases, and $\ell_2$ further divides 
the NA-CSL into the dimer and deconfining phases.
$\ell_1$ and $\ell_2$ meet at the critical point $(\epsilon_3,\ell_1(\epsilon_3))$.
Figure~\ref{fig:CSL}(d) shows 
the phase diagram in the $\epsilon$-$S$ plane
obtained by the relation $\ell(S)$;
it is useful to observe the
region around the curve corresponding to the critical velocity.
The phase boundary between the QCD vacuum
and the CSL phase represents the union of the bottom edges of the colored regions,
where the two tricritical points $\epsilon = \epsilon_{1, 2}$ can be found.
The critical velocity $S_{\rm c}$ for 
the NA-CSLs
is lower than the $S^\eta $ 
value of the $\eta$-CSL 
for $\epsilon > \epsilon_1$;
thus, it has higher potential for the discovery of a CSL in nature 
as compared to the $\eta$-CSL in Ref.~\cite{Nishimura:2020odq}.
The dimer phase is found to meet two phases among the
deconfining/confining phases and the QCD vacuum
at three tricritical points $\epsilon_{1,2,3}$.

{\it Isospinons.}  
Here, the gapless NG modes propagating 
along the lattice called isospinons (analogous to magnons
propagating along an antiferromagnetic spin chain) are discussed. 
Either in the dimer or deconfining phase,
u- and d-CSs appear sequentially, 
where the isospin moduli of the former (latter) are directed to the north (south) pole of $S^2$. 
Therefore, the isospins of neighboring NA-CSLs are antiparallel, and they are thus regarded as an antiferro-isospin chain.
There exist massless NG modes (isospinons) associated 
with $SU(2)_{\rm V} \to U(1)$  
propagating along the lattice. 
They can be found in small perturbations around the background NA-CSL:
$\phi_a = \bar\phi_a + \delta\phi_a$ ($a = 0, 1, 2, 3$), where $\bar\phi_a$ represents the background solution.

For simplicity,
the point $\beta = \epsilon = 0$
for which there is an available analytical solution is considered here: 
$\bar \phi_0 \pm \bar\phi_3 = 2\, {\rm am}\left(\frac{\zeta\mp d/2}{k},k\right) + \pi
$ and $\bar \phi_{1, 2} = 0$ 
with $d$ corresponding to the separation between u- and d-CSs.\footnote{$d$ is also a gapless mode called  
a quasi-NG mode, however, this is peculiar for $\beta = \epsilon = 0$.
In general, $d$ is gapful.}
In addition to the translational NG mode (phonon), $\delta\phi_{0,3}^{(0)} \propto \bar\phi_{0,3}'$,
there exist two gapless isospinons,
$\delta\phi_1^{(0)} \pm i\delta\phi_2^{(0)} 
\propto e^{\mp i\bar\phi_3}\bar\phi_0'$, which propagate along the u- and d-CSLs.\footnote{See 
Appendix \ref{app:b}
for the derivations of the linearized EOMs for the NA-CSL and the gapless modes.}
The phonon and isospinon robustly exist because they are NG;
they exist in NA-CSLs with generic $\beta$ and $\epsilon$ values. 
In contrast, there is only one gapless mode, a phonon, 
on the $\eta$-CSL.

{\it Rotation-induced ferro/ferrimagnetism.} 
Finally, magnetizations
appearing through another topological term
under an external magnetic field $\bm{B}$ are shown: \cite{Son:2004tq,Son:2007ny,Eto:2013hoa}: 
$
{\cal L}_{\rm top} 
=
\frac{q_{\rm u} \mu_{\rm B}}{4\pi^2}
\nabla \phi_+ \cdot \bm{B}+
\frac{q_{\rm d} \mu_{\rm B}}{4\pi^2}\nabla \phi_-
\cdot \bm{B}.
$
From this, the
magnetizations of the u-NA-CSs, d-NA-CSs, and $\eta$-CS are found as
\be
\bm{M}_{\rm u,d} = -\frac{q_{\rm u,d} \mu_{\rm B}}{4\pi^2} 
\nabla \phi_{+,-}\,,\quad
\bm{M}_{\eta} =  -\frac{1}{3}\frac{e \mu_{\rm B}}{4\pi^2} \nabla \phi_0,
\ee
respectively.
The electric charges of
up-and-down quarks  
$q_{\rm u} = 2e/3$ and $q_{\rm d} = -e/3$ 
determine the magnetizations 
$\bm{M}_{\rm u}$ and $\bm{M}_{\rm d}$ for the NA-CSLs,  
which are anti-parallel 
with different magnitudes 
and have a net magnetization, referred to as ferrimagnetism.
For the $\eta$-CSL, the magnetizations become parallel as $\bm{M}_{\eta}$,
implying ferromagnetism. 
Hence, $S_1$ ($\ell_1$) in Fig.~\ref{fig:CSL}(c) [(d)] is a critical velocity (lattice size)
separating a ferrimagnetic and ferromagnetic magnetizations.
Because rotation induces magnetization, 
this is a type of inverse gyromagnetic effect, called the Barnett effect.
This magnetism of the $\eta$-CSL is specific to 
two-flavor quarks and does not exist in three-flavor quarks, 
\cite{Nishimura:2020odq} where
the $U(1)_{\rm A}$ meson is called $\eta'$ 
(see also \cite{Huang:2017pqe}).

{\it Concluding remark.}
 The largest vorticity of the current experiment has a magnitude of the order of $\sim 10^{22}$/s \cite{STAR:2017ckg,STAR:2018gyt}.
 Although the critical rotation velocity of the $\eta$-CSL is 
 larger by one order of magnitude 
 \cite{Nishimura:2020odq}, that of
the NA-CSL is smaller than that of the $\eta$-CSL 
in all parametric regions,  
as demonstrated by this study. Thus, the NA-CSL has the potential to be reached 
in future low-energy heavy ion collision experiments.

{\it Acknowledgments}.

This work is supported in part by the JSPS Grant-in-Aid for Scientific Research 
(KAKENHI Grant
 No.~JP19K03839 (M.~E.) and 
 No.~JP18H01217 (M.~N.)). 
It is also supported by the MEXT KAKENHI Grant-in-Aid for Scientific Research on Innovative Areas
``Discrete Geometric Analysis for Materials Design'' No.~JP17H06462 (M.~E.)
 from the MEXT of Japan.
 K. N. is supported by JSPS KAKENHI Grant No. 19J21593.

\newcommand{\J}[4]{{\sl #1} {\bf #2} (#3) #4}
\newcommand{\andJ}[3]{{\bf #1} (#2) #3}
\newcommand{\AP}{Ann.\ Phys.\ (N.Y.)}
\newcommand{\MPL}{Mod.\ Phys.\ Lett.}
\newcommand{\NP}{Nucl.\ Phys.}
\newcommand{\PL}{Phys.\ Lett.}
\newcommand{\PR}{ Phys.\ Rev.}
\newcommand{\PRL}{Phys.\ Rev.\ Lett.}
\newcommand{\PTP}{Prog.\ Theor.\ Phys.}
\newcommand{\hep}[1]{{\tt hep-th/{#1}}}

\bibliography{reference.bib}

\begin{thebibliography}{59}%
\makeatletter
\providecommand \@ifxundefined [1]{%
 \@ifx{#1\undefined}
}%
\providecommand \@ifnum [1]{%
 \ifnum #1\expandafter \@firstoftwo
 \else \expandafter \@secondoftwo
 \fi
}%
\providecommand \@ifx [1]{%
 \ifx #1\expandafter \@firstoftwo
 \else \expandafter \@secondoftwo
 \fi
}%
\providecommand \natexlab [1]{#1}%
\providecommand \enquote  [1]{``#1''}%
\providecommand \bibnamefont  [1]{#1}%
\providecommand \bibfnamefont [1]{#1}%
\providecommand \citenamefont [1]{#1}%
\providecommand \href@noop [0]{\@secondoftwo}%
\providecommand \href [0]{\begingroup \@sanitize@url \@href}%
\providecommand \@href[1]{\@@startlink{#1}\@@href}%
\providecommand \@@href[1]{\endgroup#1\@@endlink}%
\providecommand \@sanitize@url [0]{\catcode `\\12\catcode `\$12\catcode
  `\&12\catcode `\#12\catcode `\^12\catcode `\_12\catcode `\%12\relax}%
\providecommand \@@startlink[1]{}%
\providecommand \@@endlink[0]{}%
\providecommand \url  [0]{\begingroup\@sanitize@url \@url }%
\providecommand \@url [1]{\endgroup\@href {#1}{\urlprefix }}%
\providecommand \urlprefix  [0]{URL }%
\providecommand \Eprint [0]{\href }%
\providecommand \doibase [0]{http://dx.doi.org/}%
\providecommand \selectlanguage [0]{\@gobble}%
\providecommand \bibinfo  [0]{\@secondoftwo}%
\providecommand \bibfield  [0]{\@secondoftwo}%
\providecommand \translation [1]{[#1]}%
\providecommand \BibitemOpen [0]{}%
\providecommand \bibitemStop [0]{}%
\providecommand \bibitemNoStop [0]{.\EOS\space}%
\providecommand \EOS [0]{\spacefactor3000\relax}%
\providecommand \BibitemShut  [1]{\csname bibitem#1\endcsname}%
\let\auto@bib@innerbib\@empty
\bibitem [{\citenamefont {Adamczyk}\ \emph {et~al.}(2017)\citenamefont
  {Adamczyk} \emph {et~al.}}]{STAR:2017ckg}%
  \BibitemOpen
  \bibfield  {author} {\bibinfo {author} {\bibfnamefont {L.}~\bibnamefont
  {Adamczyk}} \emph {et~al.} (\bibinfo {collaboration} {STAR}),\ }\href
  {\doibase 10.1038/nature23004} {\bibfield  {journal} {\bibinfo  {journal}
  {Nature}\ }\textbf {\bibinfo {volume} {548}},\ \bibinfo {pages} {62}
  (\bibinfo {year} {2017})},\ \Eprint {http://arxiv.org/abs/1701.06657}
  {arXiv:1701.06657 [nucl-ex]} \BibitemShut {NoStop}%
\bibitem [{\citenamefont {Adam}\ \emph {et~al.}(2018)\citenamefont {Adam} \emph
  {et~al.}}]{STAR:2018gyt}%
  \BibitemOpen
  \bibfield  {author} {\bibinfo {author} {\bibfnamefont {J.}~\bibnamefont
  {Adam}} \emph {et~al.} (\bibinfo {collaboration} {STAR}),\ }\href {\doibase
  10.1103/PhysRevC.98.014910} {\bibfield  {journal} {\bibinfo  {journal} {Phys.
  Rev. C}\ }\textbf {\bibinfo {volume} {98}},\ \bibinfo {pages} {014910}
  (\bibinfo {year} {2018})},\ \Eprint {http://arxiv.org/abs/1805.04400}
  {arXiv:1805.04400 [nucl-ex]} \BibitemShut {NoStop}%
\bibitem [{\citenamefont {Abbott}\ \emph {et~al.}(2017)\citenamefont {Abbott}
  \emph {et~al.}}]{TheLIGOScientific:2017qsa}%
  \BibitemOpen
  \bibfield  {author} {\bibinfo {author} {\bibfnamefont {B.~P.}\ \bibnamefont
  {Abbott}} \emph {et~al.} (\bibinfo {collaboration} {Virgo, LIGO
  Scientific}),\ }\href {\doibase 10.1103/PhysRevLett.119.161101} {\bibfield
  {journal} {\bibinfo  {journal} {Phys. Rev. Lett.}\ }\textbf {\bibinfo
  {volume} {119}},\ \bibinfo {pages} {161101} (\bibinfo {year}
  {2017})}\BibitemShut {NoStop}%
\bibitem [{\citenamefont {Abbott}\ \emph {et~al.}(2020)\citenamefont {Abbott}
  \emph {et~al.}}]{Abbott:2020uma}%
  \BibitemOpen
  \bibfield  {author} {\bibinfo {author} {\bibfnamefont {B.}~\bibnamefont
  {Abbott}} \emph {et~al.} (\bibinfo {collaboration} {LIGO Scientific,
  Virgo}),\ }\href {\doibase 10.3847/2041-8213/ab75f5} {\bibfield  {journal}
  {\bibinfo  {journal} {Astrophys. J. Lett.}\ }\textbf {\bibinfo {volume}
  {892}},\ \bibinfo {pages} {L3} (\bibinfo {year} {2020})},\ \Eprint
  {http://arxiv.org/abs/2001.01761} {arXiv:2001.01761 [astro-ph.HE]}
  \BibitemShut {NoStop}%
\bibitem [{\citenamefont {Riley}\ \emph {et~al.}(2019)\citenamefont {Riley}
  \emph {et~al.}}]{Riley:2019yda}%
  \BibitemOpen
  \bibfield  {author} {\bibinfo {author} {\bibfnamefont {T.~E.}\ \bibnamefont
  {Riley}} \emph {et~al.},\ }\href {\doibase 10.3847/2041-8213/ab481c}
  {\bibfield  {journal} {\bibinfo  {journal} {Astrophys. J. Lett.}\ }\textbf
  {\bibinfo {volume} {887}},\ \bibinfo {pages} {L21} (\bibinfo {year}
  {2019})},\ \Eprint {http://arxiv.org/abs/1912.05702} {arXiv:1912.05702
  [astro-ph.HE]} \BibitemShut {NoStop}%
\bibitem [{\citenamefont {Miller}\ \emph {et~al.}(2019)\citenamefont {Miller}
  \emph {et~al.}}]{Miller:2019cac}%
  \BibitemOpen
  \bibfield  {author} {\bibinfo {author} {\bibfnamefont {M.}~\bibnamefont
  {Miller}} \emph {et~al.},\ }\href {\doibase 10.3847/2041-8213/ab50c5}
  {\bibfield  {journal} {\bibinfo  {journal} {Astrophys. J. Lett.}\ }\textbf
  {\bibinfo {volume} {887}},\ \bibinfo {pages} {L24} (\bibinfo {year}
  {2019})},\ \Eprint {http://arxiv.org/abs/1912.05705} {arXiv:1912.05705
  [astro-ph.HE]} \BibitemShut {NoStop}%
\bibitem [{\citenamefont {Chen}\ \emph {et~al.}(2016)\citenamefont {Chen},
  \citenamefont {Fukushima}, \citenamefont {Huang},\ and\ \citenamefont
  {Mameda}}]{Chen:2015hfc}%
  \BibitemOpen
  \bibfield  {author} {\bibinfo {author} {\bibfnamefont {H.-L.}\ \bibnamefont
  {Chen}}, \bibinfo {author} {\bibfnamefont {K.}~\bibnamefont {Fukushima}},
  \bibinfo {author} {\bibfnamefont {X.-G.}\ \bibnamefont {Huang}}, \ and\
  \bibinfo {author} {\bibfnamefont {K.}~\bibnamefont {Mameda}},\ }\href
  {\doibase 10.1103/PhysRevD.93.104052} {\bibfield  {journal} {\bibinfo
  {journal} {Phys. Rev.}\ }\textbf {\bibinfo {volume} {D93}},\ \bibinfo {pages}
  {104052} (\bibinfo {year} {2016})},\ \Eprint
  {http://arxiv.org/abs/1512.08974} {arXiv:1512.08974 [hep-ph]} \BibitemShut
  {NoStop}%
\bibitem [{\citenamefont {Ebihara}\ \emph {et~al.}(2017)\citenamefont
  {Ebihara}, \citenamefont {Fukushima},\ and\ \citenamefont
  {Mameda}}]{Ebihara:2016fwa}%
  \BibitemOpen
  \bibfield  {author} {\bibinfo {author} {\bibfnamefont {S.}~\bibnamefont
  {Ebihara}}, \bibinfo {author} {\bibfnamefont {K.}~\bibnamefont {Fukushima}},
  \ and\ \bibinfo {author} {\bibfnamefont {K.}~\bibnamefont {Mameda}},\ }\href
  {\doibase 10.1016/j.physletb.2016.11.010} {\bibfield  {journal} {\bibinfo
  {journal} {Phys. Lett.}\ }\textbf {\bibinfo {volume} {B764}},\ \bibinfo
  {pages} {94} (\bibinfo {year} {2017})},\ \Eprint
  {http://arxiv.org/abs/1608.00336} {arXiv:1608.00336 [hep-ph]} \BibitemShut
  {NoStop}%
\bibitem [{\citenamefont {Jiang}\ and\ \citenamefont
  {Liao}(2016)}]{Jiang:2016wvv}%
  \BibitemOpen
  \bibfield  {author} {\bibinfo {author} {\bibfnamefont {Y.}~\bibnamefont
  {Jiang}}\ and\ \bibinfo {author} {\bibfnamefont {J.}~\bibnamefont {Liao}},\
  }\href {\doibase 10.1103/PhysRevLett.117.192302} {\bibfield  {journal}
  {\bibinfo  {journal} {Phys. Rev. Lett.}\ }\textbf {\bibinfo {volume} {117}},\
  \bibinfo {pages} {192302} (\bibinfo {year} {2016})},\ \Eprint
  {http://arxiv.org/abs/1606.03808} {arXiv:1606.03808 [hep-ph]} \BibitemShut
  {NoStop}%
\bibitem [{\citenamefont {Chernodub}\ and\ \citenamefont
  {Gongyo}(2017{\natexlab{a}})}]{Chernodub:2016kxh}%
  \BibitemOpen
  \bibfield  {author} {\bibinfo {author} {\bibfnamefont {M.~N.}\ \bibnamefont
  {Chernodub}}\ and\ \bibinfo {author} {\bibfnamefont {S.}~\bibnamefont
  {Gongyo}},\ }\href {\doibase 10.1007/JHEP01(2017)136} {\bibfield  {journal}
  {\bibinfo  {journal} {JHEP}\ }\textbf {\bibinfo {volume} {01}},\ \bibinfo
  {pages} {136} (\bibinfo {year} {2017}{\natexlab{a}})},\ \Eprint
  {http://arxiv.org/abs/1611.02598} {arXiv:1611.02598 [hep-th]} \BibitemShut
  {NoStop}%
\bibitem [{\citenamefont {Chernodub}\ and\ \citenamefont
  {Gongyo}(2017{\natexlab{b}})}]{Chernodub:2017ref}%
  \BibitemOpen
  \bibfield  {author} {\bibinfo {author} {\bibfnamefont {M.~N.}\ \bibnamefont
  {Chernodub}}\ and\ \bibinfo {author} {\bibfnamefont {S.}~\bibnamefont
  {Gongyo}},\ }\href {\doibase 10.1103/PhysRevD.95.096006} {\bibfield
  {journal} {\bibinfo  {journal} {Phys. Rev.}\ }\textbf {\bibinfo {volume}
  {D95}},\ \bibinfo {pages} {096006} (\bibinfo {year} {2017}{\natexlab{b}})},\
  \Eprint {http://arxiv.org/abs/1702.08266} {arXiv:1702.08266 [hep-th]}
  \BibitemShut {NoStop}%
\bibitem [{\citenamefont {Liu}\ and\ \citenamefont
  {Zahed}(2018)}]{Liu:2017zhl}%
  \BibitemOpen
  \bibfield  {author} {\bibinfo {author} {\bibfnamefont {Y.}~\bibnamefont
  {Liu}}\ and\ \bibinfo {author} {\bibfnamefont {I.}~\bibnamefont {Zahed}},\
  }\href {\doibase 10.1103/PhysRevD.98.014017} {\bibfield  {journal} {\bibinfo
  {journal} {Phys. Rev. D}\ }\textbf {\bibinfo {volume} {98}},\ \bibinfo
  {pages} {014017} (\bibinfo {year} {2018})},\ \Eprint
  {http://arxiv.org/abs/1710.02895} {arXiv:1710.02895 [hep-ph]} \BibitemShut
  {NoStop}%
\bibitem [{\citenamefont {Zhang}\ \emph {et~al.}(2020)\citenamefont {Zhang},
  \citenamefont {Hou},\ and\ \citenamefont {Liao}}]{Zhang:2018ome}%
  \BibitemOpen
  \bibfield  {author} {\bibinfo {author} {\bibfnamefont {H.}~\bibnamefont
  {Zhang}}, \bibinfo {author} {\bibfnamefont {D.}~\bibnamefont {Hou}}, \ and\
  \bibinfo {author} {\bibfnamefont {J.}~\bibnamefont {Liao}},\ }\href {\doibase
  10.1088/1674-1137/abae4d} {\bibfield  {journal} {\bibinfo  {journal} {Chin.
  Phys. C}\ }\textbf {\bibinfo {volume} {44}},\ \bibinfo {pages} {111001}
  (\bibinfo {year} {2020})},\ \Eprint {http://arxiv.org/abs/1812.11787}
  {arXiv:1812.11787 [hep-ph]} \BibitemShut {NoStop}%
\bibitem [{\citenamefont {Wang}\ \emph {et~al.}(2019)\citenamefont {Wang},
  \citenamefont {Jiang}, \citenamefont {He},\ and\ \citenamefont
  {Zhuang}}]{Wang:2018zrn}%
  \BibitemOpen
  \bibfield  {author} {\bibinfo {author} {\bibfnamefont {L.}~\bibnamefont
  {Wang}}, \bibinfo {author} {\bibfnamefont {Y.}~\bibnamefont {Jiang}},
  \bibinfo {author} {\bibfnamefont {L.}~\bibnamefont {He}}, \ and\ \bibinfo
  {author} {\bibfnamefont {P.}~\bibnamefont {Zhuang}},\ }\href {\doibase
  10.1103/PhysRevC.100.034902} {\bibfield  {journal} {\bibinfo  {journal}
  {Phys. Rev. C}\ }\textbf {\bibinfo {volume} {100}},\ \bibinfo {pages}
  {034902} (\bibinfo {year} {2019})},\ \Eprint
  {http://arxiv.org/abs/1901.00804} {arXiv:1901.00804 [nucl-th]} \BibitemShut
  {NoStop}%
\bibitem [{\citenamefont {Chen}\ \emph {et~al.}(2019)\citenamefont {Chen},
  \citenamefont {Huang},\ and\ \citenamefont {Mameda}}]{Chen:2019tcp}%
  \BibitemOpen
  \bibfield  {author} {\bibinfo {author} {\bibfnamefont {H.-L.}\ \bibnamefont
  {Chen}}, \bibinfo {author} {\bibfnamefont {X.-G.}\ \bibnamefont {Huang}}, \
  and\ \bibinfo {author} {\bibfnamefont {K.}~\bibnamefont {Mameda}},\
  }\href@noop {} {\  (\bibinfo {year} {2019})},\ \Eprint
  {http://arxiv.org/abs/1910.02700} {arXiv:1910.02700 [nucl-th]} \BibitemShut
  {NoStop}%
\bibitem [{\citenamefont {Chen}\ \emph
  {et~al.}(2021{\natexlab{a}})\citenamefont {Chen}, \citenamefont {Huang},\
  and\ \citenamefont {Liao}}]{Chen:2021aiq}%
  \BibitemOpen
  \bibfield  {author} {\bibinfo {author} {\bibfnamefont {H.-L.}\ \bibnamefont
  {Chen}}, \bibinfo {author} {\bibfnamefont {X.-G.}\ \bibnamefont {Huang}}, \
  and\ \bibinfo {author} {\bibfnamefont {J.}~\bibnamefont {Liao}},\ }\href
  {\doibase 10.1007/978-3-030-71427-7_11} {\bibfield  {journal} {\bibinfo
  {journal} {Lect. Notes Phys.}\ }\textbf {\bibinfo {volume} {987}},\ \bibinfo
  {pages} {349} (\bibinfo {year} {2021}{\natexlab{a}})},\ \Eprint
  {http://arxiv.org/abs/2108.00586} {arXiv:2108.00586 [hep-ph]} \BibitemShut
  {NoStop}%
\bibitem [{\citenamefont {Huang}\ \emph {et~al.}(2018)\citenamefont {Huang},
  \citenamefont {Nishimura},\ and\ \citenamefont {Yamamoto}}]{Huang:2017pqe}%
  \BibitemOpen
  \bibfield  {author} {\bibinfo {author} {\bibfnamefont {X.-G.}\ \bibnamefont
  {Huang}}, \bibinfo {author} {\bibfnamefont {K.}~\bibnamefont {Nishimura}}, \
  and\ \bibinfo {author} {\bibfnamefont {N.}~\bibnamefont {Yamamoto}},\ }\href
  {\doibase 10.1007/JHEP02(2018)069} {\bibfield  {journal} {\bibinfo  {journal}
  {JHEP}\ }\textbf {\bibinfo {volume} {02}},\ \bibinfo {pages} {069} (\bibinfo
  {year} {2018})},\ \Eprint {http://arxiv.org/abs/1711.02190} {arXiv:1711.02190
  [hep-ph]} \BibitemShut {NoStop}%
\bibitem [{\citenamefont {Nishimura}\ and\ \citenamefont
  {Yamamoto}(2020)}]{Nishimura:2020odq}%
  \BibitemOpen
  \bibfield  {author} {\bibinfo {author} {\bibfnamefont {K.}~\bibnamefont
  {Nishimura}}\ and\ \bibinfo {author} {\bibfnamefont {N.}~\bibnamefont
  {Yamamoto}},\ }\href {\doibase 10.1007/JHEP07(2020)196} {\bibfield  {journal}
  {\bibinfo  {journal} {JHEP}\ }\textbf {\bibinfo {volume} {07}},\ \bibinfo
  {pages} {196} (\bibinfo {year} {2020})},\ \Eprint
  {http://arxiv.org/abs/2003.13945} {arXiv:2003.13945 [hep-ph]} \BibitemShut
  {NoStop}%
\bibitem [{\citenamefont {Vilenkin}(1979)}]{Vilenkin:1979ui}%
  \BibitemOpen
  \bibfield  {author} {\bibinfo {author} {\bibfnamefont {A.}~\bibnamefont
  {Vilenkin}},\ }\href {\doibase 10.1103/PhysRevD.20.1807} {\bibfield
  {journal} {\bibinfo  {journal} {Phys. Rev. D}\ }\textbf {\bibinfo {volume}
  {20}},\ \bibinfo {pages} {1807} (\bibinfo {year} {1979})}\BibitemShut
  {NoStop}%
\bibitem [{\citenamefont {Vilenkin}(1980)}]{Vilenkin:1980zv}%
  \BibitemOpen
  \bibfield  {author} {\bibinfo {author} {\bibfnamefont {A.}~\bibnamefont
  {Vilenkin}},\ }\href {\doibase 10.1103/PhysRevD.21.2260} {\bibfield
  {journal} {\bibinfo  {journal} {Phys. Rev. D}\ }\textbf {\bibinfo {volume}
  {21}},\ \bibinfo {pages} {2260} (\bibinfo {year} {1980})}\BibitemShut
  {NoStop}%
\bibitem [{\citenamefont {Son}\ and\ \citenamefont
  {Surowka}(2009)}]{Son:2009tf}%
  \BibitemOpen
  \bibfield  {author} {\bibinfo {author} {\bibfnamefont {D.~T.}\ \bibnamefont
  {Son}}\ and\ \bibinfo {author} {\bibfnamefont {P.}~\bibnamefont {Surowka}},\
  }\href {\doibase 10.1103/PhysRevLett.103.191601} {\bibfield  {journal}
  {\bibinfo  {journal} {Phys. Rev. Lett.}\ }\textbf {\bibinfo {volume} {103}},\
  \bibinfo {pages} {191601} (\bibinfo {year} {2009})},\ \Eprint
  {http://arxiv.org/abs/0906.5044} {arXiv:0906.5044 [hep-th]} \BibitemShut
  {NoStop}%
\bibitem [{\citenamefont {Landsteiner}\ \emph {et~al.}(2011)\citenamefont
  {Landsteiner}, \citenamefont {Megias},\ and\ \citenamefont
  {Pena-Benitez}}]{Landsteiner:2011cp}%
  \BibitemOpen
  \bibfield  {author} {\bibinfo {author} {\bibfnamefont {K.}~\bibnamefont
  {Landsteiner}}, \bibinfo {author} {\bibfnamefont {E.}~\bibnamefont {Megias}},
  \ and\ \bibinfo {author} {\bibfnamefont {F.}~\bibnamefont {Pena-Benitez}},\
  }\href {\doibase 10.1103/PhysRevLett.107.021601} {\bibfield  {journal}
  {\bibinfo  {journal} {Phys. Rev. Lett.}\ }\textbf {\bibinfo {volume} {107}},\
  \bibinfo {pages} {021601} (\bibinfo {year} {2011})},\ \Eprint
  {http://arxiv.org/abs/1103.5006} {arXiv:1103.5006 [hep-ph]} \BibitemShut
  {NoStop}%
\bibitem [{\citenamefont {Landsteiner}\ \emph {et~al.}(2013)\citenamefont
  {Landsteiner}, \citenamefont {Megias},\ and\ \citenamefont
  {Pena-Benitez}}]{Landsteiner:2012kd}%
  \BibitemOpen
  \bibfield  {author} {\bibinfo {author} {\bibfnamefont {K.}~\bibnamefont
  {Landsteiner}}, \bibinfo {author} {\bibfnamefont {E.}~\bibnamefont {Megias}},
  \ and\ \bibinfo {author} {\bibfnamefont {F.}~\bibnamefont {Pena-Benitez}},\
  }\href {\doibase 10.1007/978-3-642-37305-3_17} {\bibfield  {journal}
  {\bibinfo  {journal} {Lect. Notes Phys.}\ }\textbf {\bibinfo {volume}
  {871}},\ \bibinfo {pages} {433} (\bibinfo {year} {2013})},\ \Eprint
  {http://arxiv.org/abs/1207.5808} {arXiv:1207.5808 [hep-th]} \BibitemShut
  {NoStop}%
\bibitem [{\citenamefont {Landsteiner}(2016)}]{Landsteiner:2016led}%
  \BibitemOpen
  \bibfield  {author} {\bibinfo {author} {\bibfnamefont {K.}~\bibnamefont
  {Landsteiner}},\ }\href {\doibase 10.5506/APhysPolB.47.2617} {\bibfield
  {journal} {\bibinfo  {journal} {Acta Phys. Polon. B}\ }\textbf {\bibinfo
  {volume} {47}},\ \bibinfo {pages} {2617} (\bibinfo {year} {2016})},\ \Eprint
  {http://arxiv.org/abs/1610.04413} {arXiv:1610.04413 [hep-th]} \BibitemShut
  {NoStop}%
\bibitem [{\citenamefont {Son}\ and\ \citenamefont
  {Stephanov}(2008)}]{Son:2007ny}%
  \BibitemOpen
  \bibfield  {author} {\bibinfo {author} {\bibfnamefont {D.~T.}\ \bibnamefont
  {Son}}\ and\ \bibinfo {author} {\bibfnamefont {M.~A.}\ \bibnamefont
  {Stephanov}},\ }\href {\doibase 10.1103/PhysRevD.77.014021} {\bibfield
  {journal} {\bibinfo  {journal} {Phys. Rev. D}\ }\textbf {\bibinfo {volume}
  {77}},\ \bibinfo {pages} {014021} (\bibinfo {year} {2008})},\ \Eprint
  {http://arxiv.org/abs/0710.1084} {arXiv:0710.1084 [hep-ph]} \BibitemShut
  {NoStop}%
\bibitem [{\citenamefont {Eto}\ \emph {et~al.}(2013)\citenamefont {Eto},
  \citenamefont {Hashimoto},\ and\ \citenamefont {Hatsuda}}]{Eto:2012qd}%
  \BibitemOpen
  \bibfield  {author} {\bibinfo {author} {\bibfnamefont {M.}~\bibnamefont
  {Eto}}, \bibinfo {author} {\bibfnamefont {K.}~\bibnamefont {Hashimoto}}, \
  and\ \bibinfo {author} {\bibfnamefont {T.}~\bibnamefont {Hatsuda}},\ }\href
  {\doibase 10.1103/PhysRevD.88.081701} {\bibfield  {journal} {\bibinfo
  {journal} {Phys. Rev. D}\ }\textbf {\bibinfo {volume} {88}},\ \bibinfo
  {pages} {081701} (\bibinfo {year} {2013})},\ \Eprint
  {http://arxiv.org/abs/1209.4814} {arXiv:1209.4814 [hep-ph]} \BibitemShut
  {NoStop}%
\bibitem [{\citenamefont {Brauner}\ and\ \citenamefont
  {Yamamoto}(2017)}]{Brauner:2016pko}%
  \BibitemOpen
  \bibfield  {author} {\bibinfo {author} {\bibfnamefont {T.}~\bibnamefont
  {Brauner}}\ and\ \bibinfo {author} {\bibfnamefont {N.}~\bibnamefont
  {Yamamoto}},\ }\href {\doibase 10.1007/JHEP04(2017)132} {\bibfield  {journal}
  {\bibinfo  {journal} {JHEP}\ }\textbf {\bibinfo {volume} {04}},\ \bibinfo
  {pages} {132} (\bibinfo {year} {2017})},\ \Eprint
  {http://arxiv.org/abs/1609.05213} {arXiv:1609.05213 [hep-ph]} \BibitemShut
  {NoStop}%
\bibitem [{\citenamefont {Chen}\ \emph
  {et~al.}(2021{\natexlab{b}})\citenamefont {Chen}, \citenamefont {Fukushima},\
  and\ \citenamefont {Qiu}}]{Chen:2021vou}%
  \BibitemOpen
  \bibfield  {author} {\bibinfo {author} {\bibfnamefont {S.}~\bibnamefont
  {Chen}}, \bibinfo {author} {\bibfnamefont {K.}~\bibnamefont {Fukushima}}, \
  and\ \bibinfo {author} {\bibfnamefont {Z.}~\bibnamefont {Qiu}},\ }\href@noop
  {} {\  (\bibinfo {year} {2021}{\natexlab{b}})},\ \Eprint
  {http://arxiv.org/abs/2104.11482} {arXiv:2104.11482 [hep-ph]} \BibitemShut
  {NoStop}%
\bibitem [{\citenamefont {Brauner}\ and\ \citenamefont
  {Kadam}(2017{\natexlab{a}})}]{Brauner:2017uiu}%
  \BibitemOpen
  \bibfield  {author} {\bibinfo {author} {\bibfnamefont {T.}~\bibnamefont
  {Brauner}}\ and\ \bibinfo {author} {\bibfnamefont {S.~V.}\ \bibnamefont
  {Kadam}},\ }\href {\doibase 10.1007/JHEP11(2017)103} {\bibfield  {journal}
  {\bibinfo  {journal} {JHEP}\ }\textbf {\bibinfo {volume} {11}},\ \bibinfo
  {pages} {103} (\bibinfo {year} {2017}{\natexlab{a}})},\ \Eprint
  {http://arxiv.org/abs/1706.04514} {arXiv:1706.04514 [hep-ph]} \BibitemShut
  {NoStop}%
\bibitem [{\citenamefont {Brauner}\ and\ \citenamefont
  {Kadam}(2017{\natexlab{b}})}]{Brauner:2017mui}%
  \BibitemOpen
  \bibfield  {author} {\bibinfo {author} {\bibfnamefont {T.}~\bibnamefont
  {Brauner}}\ and\ \bibinfo {author} {\bibfnamefont {S.}~\bibnamefont
  {Kadam}},\ }\href {\doibase 10.1007/JHEP03(2017)015} {\bibfield  {journal}
  {\bibinfo  {journal} {JHEP}\ }\textbf {\bibinfo {volume} {03}},\ \bibinfo
  {pages} {015} (\bibinfo {year} {2017}{\natexlab{b}})},\ \Eprint
  {http://arxiv.org/abs/1701.06793} {arXiv:1701.06793 [hep-ph]} \BibitemShut
  {NoStop}%
\bibitem [{\citenamefont {Brauner}\ \emph {et~al.}(2021)\citenamefont
  {Brauner}, \citenamefont {Kole\v{s}ov\'a},\ and\ \citenamefont
  {Yamamoto}}]{Brauner:2021sci}%
  \BibitemOpen
  \bibfield  {author} {\bibinfo {author} {\bibfnamefont {T.}~\bibnamefont
  {Brauner}}, \bibinfo {author} {\bibfnamefont {H.}~\bibnamefont
  {Kole\v{s}ov\'a}}, \ and\ \bibinfo {author} {\bibfnamefont {N.}~\bibnamefont
  {Yamamoto}},\ }\href {\doibase 10.1016/j.physletb.2021.136767} {\bibfield
  {journal} {\bibinfo  {journal} {Phys. Lett. B}\ }\textbf {\bibinfo {volume}
  {823}},\ \bibinfo {pages} {136767} (\bibinfo {year} {2021})},\ \Eprint
  {http://arxiv.org/abs/2108.10044} {arXiv:2108.10044 [hep-ph]} \BibitemShut
  {NoStop}%
\bibitem [{\citenamefont {Yamada}\ and\ \citenamefont
  {Yamamoto}(2021)}]{Yamada:2021jhy}%
  \BibitemOpen
  \bibfield  {author} {\bibinfo {author} {\bibfnamefont {A.}~\bibnamefont
  {Yamada}}\ and\ \bibinfo {author} {\bibfnamefont {N.}~\bibnamefont
  {Yamamoto}},\ }\href {\doibase 10.1103/PhysRevD.104.054041} {\bibfield
  {journal} {\bibinfo  {journal} {Phys. Rev. D}\ }\textbf {\bibinfo {volume}
  {104}},\ \bibinfo {pages} {054041} (\bibinfo {year} {2021})},\ \Eprint
  {http://arxiv.org/abs/2107.07074} {arXiv:2107.07074 [hep-ph]} \BibitemShut
  {NoStop}%
\bibitem [{\citenamefont {Brauner}\ \emph
  {et~al.}(2019{\natexlab{a}})\citenamefont {Brauner}, \citenamefont {Filios},\
  and\ \citenamefont {Kole\v{s}ov\'a}}]{Brauner:2019aid}%
  \BibitemOpen
  \bibfield  {author} {\bibinfo {author} {\bibfnamefont {T.}~\bibnamefont
  {Brauner}}, \bibinfo {author} {\bibfnamefont {G.}~\bibnamefont {Filios}}, \
  and\ \bibinfo {author} {\bibfnamefont {H.}~\bibnamefont {Kole\v{s}ov\'a}},\
  }\href {\doibase 10.1007/JHEP12(2019)029} {\bibfield  {journal} {\bibinfo
  {journal} {JHEP}\ }\textbf {\bibinfo {volume} {12}},\ \bibinfo {pages} {029}
  (\bibinfo {year} {2019}{\natexlab{a}})},\ \Eprint
  {http://arxiv.org/abs/1905.11409} {arXiv:1905.11409 [hep-ph]} \BibitemShut
  {NoStop}%
\bibitem [{\citenamefont {Brauner}\ \emph
  {et~al.}(2019{\natexlab{b}})\citenamefont {Brauner}, \citenamefont {Filios},\
  and\ \citenamefont {Kole\v{s}ov\'a}}]{Brauner:2019rjg}%
  \BibitemOpen
  \bibfield  {author} {\bibinfo {author} {\bibfnamefont {T.}~\bibnamefont
  {Brauner}}, \bibinfo {author} {\bibfnamefont {G.}~\bibnamefont {Filios}}, \
  and\ \bibinfo {author} {\bibfnamefont {H.}~\bibnamefont {Kole\v{s}ov\'a}},\
  }\href {\doibase 10.1103/PhysRevLett.123.012001} {\bibfield  {journal}
  {\bibinfo  {journal} {Phys. Rev. Lett.}\ }\textbf {\bibinfo {volume} {123}},\
  \bibinfo {pages} {012001} (\bibinfo {year} {2019}{\natexlab{b}})},\ \Eprint
  {http://arxiv.org/abs/1902.07522} {arXiv:1902.07522 [hep-ph]} \BibitemShut
  {NoStop}%
\bibitem [{\citenamefont {Dzyaloshinsky}(1964)}]{Dzyaloshinsky:1964dz}%
  \BibitemOpen
  \bibfield  {author} {\bibinfo {author} {\bibfnamefont {I.~E.}\ \bibnamefont
  {Dzyaloshinsky}},\ }\href@noop {} {\bibfield  {journal} {\bibinfo  {journal}
  {Sov. Phys. JETP}\ }\textbf {\bibinfo {volume} {19}},\ \bibinfo {pages} {960}
  (\bibinfo {year} {1964})}\BibitemShut {NoStop}%
\bibitem [{\citenamefont {Togawa}\ \emph {et~al.}(2012)\citenamefont {Togawa},
  \citenamefont {Koyama}, \citenamefont {Takayanagi}, \citenamefont {Mori},
  \citenamefont {Kousaka}, \citenamefont {Akimitsu}, \citenamefont {Nishihara},
  \citenamefont {Inoue}, \citenamefont {Ovchinnikov},\ and\ \citenamefont
  {Kishine}}]{togawa2012chiral}%
  \BibitemOpen
  \bibfield  {author} {\bibinfo {author} {\bibfnamefont {Y.}~\bibnamefont
  {Togawa}}, \bibinfo {author} {\bibfnamefont {T.}~\bibnamefont {Koyama}},
  \bibinfo {author} {\bibfnamefont {K.}~\bibnamefont {Takayanagi}}, \bibinfo
  {author} {\bibfnamefont {S.}~\bibnamefont {Mori}}, \bibinfo {author}
  {\bibfnamefont {Y.}~\bibnamefont {Kousaka}}, \bibinfo {author} {\bibfnamefont
  {J.}~\bibnamefont {Akimitsu}}, \bibinfo {author} {\bibfnamefont
  {S.}~\bibnamefont {Nishihara}}, \bibinfo {author} {\bibfnamefont
  {K.}~\bibnamefont {Inoue}}, \bibinfo {author} {\bibfnamefont
  {A.}~\bibnamefont {Ovchinnikov}}, \ and\ \bibinfo {author} {\bibfnamefont
  {J.-i.}\ \bibnamefont {Kishine}},\ }\href@noop {} {\bibfield  {journal}
  {\bibinfo  {journal} {Physical review letters}\ }\textbf {\bibinfo {volume}
  {108}},\ \bibinfo {pages} {107202} (\bibinfo {year} {2012})}\BibitemShut
  {NoStop}%
\bibitem [{\citenamefont {Togawa}\ \emph {et~al.}(2016)\citenamefont {Togawa},
  \citenamefont {Kousaka}, \citenamefont {Inoue},\ and\ \citenamefont
  {Kishine}}]{togawa2016symmetry}%
  \BibitemOpen
  \bibfield  {author} {\bibinfo {author} {\bibfnamefont {Y.}~\bibnamefont
  {Togawa}}, \bibinfo {author} {\bibfnamefont {Y.}~\bibnamefont {Kousaka}},
  \bibinfo {author} {\bibfnamefont {K.}~\bibnamefont {Inoue}}, \ and\ \bibinfo
  {author} {\bibfnamefont {J.-i.}\ \bibnamefont {Kishine}},\ }\href@noop {}
  {\bibfield  {journal} {\bibinfo  {journal} {Journal of the Physical Society
  of Japan}\ }\textbf {\bibinfo {volume} {85}},\ \bibinfo {pages} {112001}
  (\bibinfo {year} {2016})}\BibitemShut {NoStop}%
\bibitem [{\citenamefont {Nitta}(2015)}]{Nitta:2014rxa}%
  \BibitemOpen
  \bibfield  {author} {\bibinfo {author} {\bibfnamefont {M.}~\bibnamefont
  {Nitta}},\ }\href {\doibase 10.1016/j.nuclphysb.2015.04.006} {\bibfield
  {journal} {\bibinfo  {journal} {Nucl. Phys. B}\ }\textbf {\bibinfo {volume}
  {895}},\ \bibinfo {pages} {288} (\bibinfo {year} {2015})},\ \Eprint
  {http://arxiv.org/abs/1412.8276} {arXiv:1412.8276 [hep-th]} \BibitemShut
  {NoStop}%
\bibitem [{\citenamefont {Eto}\ and\ \citenamefont
  {Nitta}(2015)}]{Eto:2015uqa}%
  \BibitemOpen
  \bibfield  {author} {\bibinfo {author} {\bibfnamefont {M.}~\bibnamefont
  {Eto}}\ and\ \bibinfo {author} {\bibfnamefont {M.}~\bibnamefont {Nitta}},\
  }\href {\doibase 10.1103/PhysRevD.91.085044} {\bibfield  {journal} {\bibinfo
  {journal} {Phys. Rev. D}\ }\textbf {\bibinfo {volume} {91}},\ \bibinfo
  {pages} {085044} (\bibinfo {year} {2015})},\ \Eprint
  {http://arxiv.org/abs/1501.07038} {arXiv:1501.07038 [hep-th]} \BibitemShut
  {NoStop}%
\bibitem [{\citenamefont {Son}\ and\ \citenamefont
  {Zhitnitsky}(2004)}]{Son:2004tq}%
  \BibitemOpen
  \bibfield  {author} {\bibinfo {author} {\bibfnamefont {D.~T.}\ \bibnamefont
  {Son}}\ and\ \bibinfo {author} {\bibfnamefont {A.~R.}\ \bibnamefont
  {Zhitnitsky}},\ }\href {\doibase 10.1103/PhysRevD.70.074018} {\bibfield
  {journal} {\bibinfo  {journal} {Phys. Rev. D}\ }\textbf {\bibinfo {volume}
  {70}},\ \bibinfo {pages} {074018} (\bibinfo {year} {2004})},\ \Eprint
  {http://arxiv.org/abs/hep-ph/0405216} {arXiv:hep-ph/0405216} \BibitemShut
  {NoStop}%
\bibitem [{\citenamefont {Eto}\ \emph {et~al.}(2014)\citenamefont {Eto},
  \citenamefont {Hirono}, \citenamefont {Nitta},\ and\ \citenamefont
  {Yasui}}]{Eto:2013hoa}%
  \BibitemOpen
  \bibfield  {author} {\bibinfo {author} {\bibfnamefont {M.}~\bibnamefont
  {Eto}}, \bibinfo {author} {\bibfnamefont {Y.}~\bibnamefont {Hirono}},
  \bibinfo {author} {\bibfnamefont {M.}~\bibnamefont {Nitta}}, \ and\ \bibinfo
  {author} {\bibfnamefont {S.}~\bibnamefont {Yasui}},\ }\href {\doibase
  10.1093/ptep/ptt095} {\bibfield  {journal} {\bibinfo  {journal} {PTEP}\
  }\textbf {\bibinfo {volume} {2014}},\ \bibinfo {pages} {012D01} (\bibinfo
  {year} {2014})},\ \Eprint {http://arxiv.org/abs/1308.1535} {arXiv:1308.1535
  [hep-ph]} \BibitemShut {NoStop}%
\bibitem [{Note1()}]{Note1}%
  \BibitemOpen
  \bibinfo {note} {See Appendix \ref {app:a} for details on the CVE
  term.}\BibitemShut {Stop}%
\bibitem [{Note2()}]{Note2}%
  \BibitemOpen
  \bibinfo {note} {See Appendix \ref {app:a} for the potential and vacuum with
  various parameter choices.}\BibitemShut {Stop}%
\bibitem [{Note3()}]{Note3}%
  \BibitemOpen
  \bibinfo {note} {See Appendix \ref {app:b} for the derivations of the
  linearized EOMs for the $\eta $-CSL.}\BibitemShut {Stop}%
\bibitem [{Note4()}]{Note4}%
  \BibitemOpen
  \bibinfo {note} {In the three-flavor case \cite {Nishimura:2020odq}, the
  ratio $f_{\eta '}/f_\pi $ for the vacuum values is estimated as $\approx
  1.1$, implying that $\epsilon \approx 0.17 > 0$.}\BibitemShut {Stop}%
\bibitem [{Note5()}]{Note5}%
  \BibitemOpen
  \bibinfo {note} {See Appendix \ref {app:a} for the derivation of the
  EOMs.}\BibitemShut {Stop}%
\bibitem [{Note6()}]{Note6}%
  \BibitemOpen
  \bibinfo {note} {See Appendix \ref {app:a} for the numerical solutions with
  various parameters.}\BibitemShut {Stop}%
\bibitem [{Note7()}]{Note7}%
  \BibitemOpen
  \bibinfo {note} {$d$ is also a gapless mode called a quasi-NG mode, however,
  this is peculiar for $\beta = \epsilon = 0$. In general, $d$ is
  gapful.}\BibitemShut {Stop}%
\bibitem [{Note8()}]{Note8}%
  \BibitemOpen
  \bibinfo {note} {See Appendix \ref {app:b} for the derivations of the
  linearized EOMs for the NA-CSL and the gapless modes.}\BibitemShut {Stop}%
\bibitem [{\citenamefont {Nitta}(1999)}]{Nitta:1998qp}%
  \BibitemOpen
  \bibfield  {author} {\bibinfo {author} {\bibfnamefont {M.}~\bibnamefont
  {Nitta}},\ }\href {\doibase 10.1142/S0217751X99001202} {\bibfield  {journal}
  {\bibinfo  {journal} {Int. J. Mod. Phys. A}\ }\textbf {\bibinfo {volume}
  {14}},\ \bibinfo {pages} {2397} (\bibinfo {year} {1999})},\ \Eprint
  {http://arxiv.org/abs/hep-th/9805038} {arXiv:hep-th/9805038} \BibitemShut
  {NoStop}%
\bibitem [{\citenamefont {Nitta}\ and\ \citenamefont
  {Takahashi}(2015)}]{Nitta:2014jta}%
  \BibitemOpen
  \bibfield  {author} {\bibinfo {author} {\bibfnamefont {M.}~\bibnamefont
  {Nitta}}\ and\ \bibinfo {author} {\bibfnamefont {D.~A.}\ \bibnamefont
  {Takahashi}},\ }\href {\doibase 10.1103/PhysRevD.91.025018} {\bibfield
  {journal} {\bibinfo  {journal} {Phys. Rev. D}\ }\textbf {\bibinfo {volume}
  {91}},\ \bibinfo {pages} {025018} (\bibinfo {year} {2015})},\ \Eprint
  {http://arxiv.org/abs/1410.2391} {arXiv:1410.2391 [hep-th]} \BibitemShut
  {NoStop}%
\bibitem [{\citenamefont {Eto}\ \emph {et~al.}(2008)\citenamefont {Eto},
  \citenamefont {Fujimori}, \citenamefont {Nitta}, \citenamefont {Ohashi},\
  and\ \citenamefont {Sakai}}]{Eto:2008dm}%
  \BibitemOpen
  \bibfield  {author} {\bibinfo {author} {\bibfnamefont {M.}~\bibnamefont
  {Eto}}, \bibinfo {author} {\bibfnamefont {T.}~\bibnamefont {Fujimori}},
  \bibinfo {author} {\bibfnamefont {M.}~\bibnamefont {Nitta}}, \bibinfo
  {author} {\bibfnamefont {K.}~\bibnamefont {Ohashi}}, \ and\ \bibinfo {author}
  {\bibfnamefont {N.}~\bibnamefont {Sakai}},\ }\href {\doibase
  10.1103/PhysRevD.77.125008} {\bibfield  {journal} {\bibinfo  {journal} {Phys.
  Rev. D}\ }\textbf {\bibinfo {volume} {77}},\ \bibinfo {pages} {125008}
  (\bibinfo {year} {2008})},\ \Eprint {http://arxiv.org/abs/0802.3135}
  {arXiv:0802.3135 [hep-th]} \BibitemShut {NoStop}%
\bibitem [{\citenamefont {Condat}\ \emph {et~al.}(1983)\citenamefont {Condat},
  \citenamefont {Guyer},\ and\ \citenamefont {Miller}}]{PhysRevB.27.474}%
  \BibitemOpen
  \bibfield  {author} {\bibinfo {author} {\bibfnamefont {C.~A.}\ \bibnamefont
  {Condat}}, \bibinfo {author} {\bibfnamefont {R.~A.}\ \bibnamefont {Guyer}}, \
  and\ \bibinfo {author} {\bibfnamefont {M.~D.}\ \bibnamefont {Miller}},\
  }\href {\doibase 10.1103/PhysRevB.27.474} {\bibfield  {journal} {\bibinfo
  {journal} {Phys. Rev. B}\ }\textbf {\bibinfo {volume} {27}},\ \bibinfo
  {pages} {474} (\bibinfo {year} {1983})}\BibitemShut {NoStop}%
\bibitem [{\citenamefont {Ross}\ \emph {et~al.}(2020)\citenamefont {Ross},
  \citenamefont {Sakai},\ and\ \citenamefont {Nitta}}]{Ross:2020orc}%
  \BibitemOpen
  \bibfield  {author} {\bibinfo {author} {\bibfnamefont {C.}~\bibnamefont
  {Ross}}, \bibinfo {author} {\bibfnamefont {N.}~\bibnamefont {Sakai}}, \ and\
  \bibinfo {author} {\bibfnamefont {M.}~\bibnamefont {Nitta}},\ }\href@noop {}
  {\bibfield  {journal} {\bibinfo  {journal} {JHEP (to appear)}\ } (\bibinfo
  {year} {2020})},\ \Eprint {http://arxiv.org/abs/2012.08800} {arXiv:2012.08800
  [cond-mat.mes-hall]} \BibitemShut {NoStop}%
\bibitem [{Note9()}]{Note9}%
  \BibitemOpen
  \bibinfo {note} {Quasi-NG modes are gapless modes which are not related to
  symmetry breaking of the whole system \cite {Nitta:1998qp,Nitta:2014jta}.
  Quasi-NG modes associated to the presence of parallel domain walls were found
  in Ref.~\cite {Eto:2008dm}. In this case, the exchange between NG and
  quasi-NG modes with keeping the total number occurs \cite {Nitta:1998qp} when
  two solitons coincide. In the NA-CSL for $\beta = \epsilon = 0$, there are
  three NG modes (translation and isospinons) and one quasi-NG mode when the
  u-CSL and d-CSL are separated, while there are one NG mode (translation) and
  three quasi-NG modes when they coincide.}\BibitemShut {Stop}%
\bibitem [{\citenamefont {Watanabe}\ and\ \citenamefont
  {Murayama}(2012)}]{Watanabe:2012hr}%
  \BibitemOpen
  \bibfield  {author} {\bibinfo {author} {\bibfnamefont {H.}~\bibnamefont
  {Watanabe}}\ and\ \bibinfo {author} {\bibfnamefont {H.}~\bibnamefont
  {Murayama}},\ }\href {\doibase 10.1103/PhysRevLett.108.251602} {\bibfield
  {journal} {\bibinfo  {journal} {Phys. Rev. Lett.}\ }\textbf {\bibinfo
  {volume} {108}},\ \bibinfo {pages} {251602} (\bibinfo {year} {2012})},\
  \Eprint {http://arxiv.org/abs/1203.0609} {arXiv:1203.0609 [hep-th]}
  \BibitemShut {NoStop}%
\bibitem [{\citenamefont {Hidaka}(2013)}]{Hidaka:2012ym}%
  \BibitemOpen
  \bibfield  {author} {\bibinfo {author} {\bibfnamefont {Y.}~\bibnamefont
  {Hidaka}},\ }\href {\doibase 10.1103/PhysRevLett.110.091601} {\bibfield
  {journal} {\bibinfo  {journal} {Phys. Rev. Lett.}\ }\textbf {\bibinfo
  {volume} {110}},\ \bibinfo {pages} {091601} (\bibinfo {year} {2013})},\
  \Eprint {http://arxiv.org/abs/1203.1494} {arXiv:1203.1494 [hep-th]}
  \BibitemShut {NoStop}%
\bibitem [{\citenamefont {Takahashi}\ and\ \citenamefont
  {Nitta}(2015)}]{Takahashi:2014vua}%
  \BibitemOpen
  \bibfield  {author} {\bibinfo {author} {\bibfnamefont {D.~A.}\ \bibnamefont
  {Takahashi}}\ and\ \bibinfo {author} {\bibfnamefont {M.}~\bibnamefont
  {Nitta}},\ }\href {\doibase 10.1016/j.aop.2014.12.009} {\bibfield  {journal}
  {\bibinfo  {journal} {Annals Phys.}\ }\textbf {\bibinfo {volume} {354}},\
  \bibinfo {pages} {101} (\bibinfo {year} {2015})},\ \Eprint
  {http://arxiv.org/abs/1404.7696} {arXiv:1404.7696 [cond-mat.quant-gas]}
  \BibitemShut {NoStop}%
\bibitem [{\citenamefont {Kobayashi}\ \emph {et~al.}(2014)\citenamefont
  {Kobayashi}, \citenamefont {Nakano},\ and\ \citenamefont
  {Nitta}}]{Kobayashi:2013axa}%
  \BibitemOpen
  \bibfield  {author} {\bibinfo {author} {\bibfnamefont {M.}~\bibnamefont
  {Kobayashi}}, \bibinfo {author} {\bibfnamefont {E.}~\bibnamefont {Nakano}}, \
  and\ \bibinfo {author} {\bibfnamefont {M.}~\bibnamefont {Nitta}},\ }\href
  {\doibase 10.1007/JHEP06(2014)130} {\bibfield  {journal} {\bibinfo  {journal}
  {JHEP}\ }\textbf {\bibinfo {volume} {06}},\ \bibinfo {pages} {130} (\bibinfo
  {year} {2014})},\ \Eprint {http://arxiv.org/abs/1311.2399} {arXiv:1311.2399
  [hep-ph]} \BibitemShut {NoStop}%
\end{thebibliography}%

\onecolumngrid
\appendix

\renewcommand\thefigure{\thesection.\arabic{figure}}    
\section{Some formulae for chiral Lagrangian under rotation}
\label{app:a}
\setcounter{figure}{0}

Here, we summarize rotational effects 
on the chiral Lagrangian. 
The metric 
in the rotating frame 
around the $z$-axis 
is given by
\be
g_{\mu\nu} = \left(
\begin{array}{cccc}
1-\Omega^2(x^2+y^2) & \Omega y & -\Omega x & 0 \\
\Omega y & -1 & 0 & 0\\
-\Omega x & 0 & - 1 & 0\\
0 & 0 & 0 & -1
\end{array}
\right),
\label{eq:metric_rotation}
\ee
where $\Omega$ is an angular velocity.
In addition, the rotation effect appears via 
the chiral vortical effect (CVE) which is represented in the chiral Lagrangian
\cite{Huang:2017pqe,Nishimura:2020odq} by
\be
{\cal L}_{\rm CVE} = 
\frac{N_c}{4\pi^2}\mu_b\mu_c\left(
d_{Abc}\,{\bf \Omega} \cdot \bm{\nabla} \frac{\pi_A}{f_\pi}
+ 
d_{0bc}\,{\bf \Omega} \cdot \bm{\nabla} \frac{\eta}{f_\eta}\right)
= \frac{\Omega}{2\pi^2N_c}\mu_B^2\p_z\frac{\eta}{f_\eta}
\label{eq:CVE}
\ee
with $d_{abc} = \frac{1}{2}\,{\rm Tr}\left[\lambda_a\left\{\lambda_b,\lambda_c\right\}\right]$
(a = 0, 1, 2, 3), and at the rightmost equality, $\mu_A = 0$ $(A = 1, 2, 3)$, and 
the zeroth component $\mu_0 = \mu_{\rm B}$, which is valid for baryonic matter, are set.
The CVE term in Eq.~(\ref{eq:CVE}) is used in Eq.~(1) of the the main text.

\begin{figure}[h]
\begin{center}
\includegraphics[width=\textwidth]{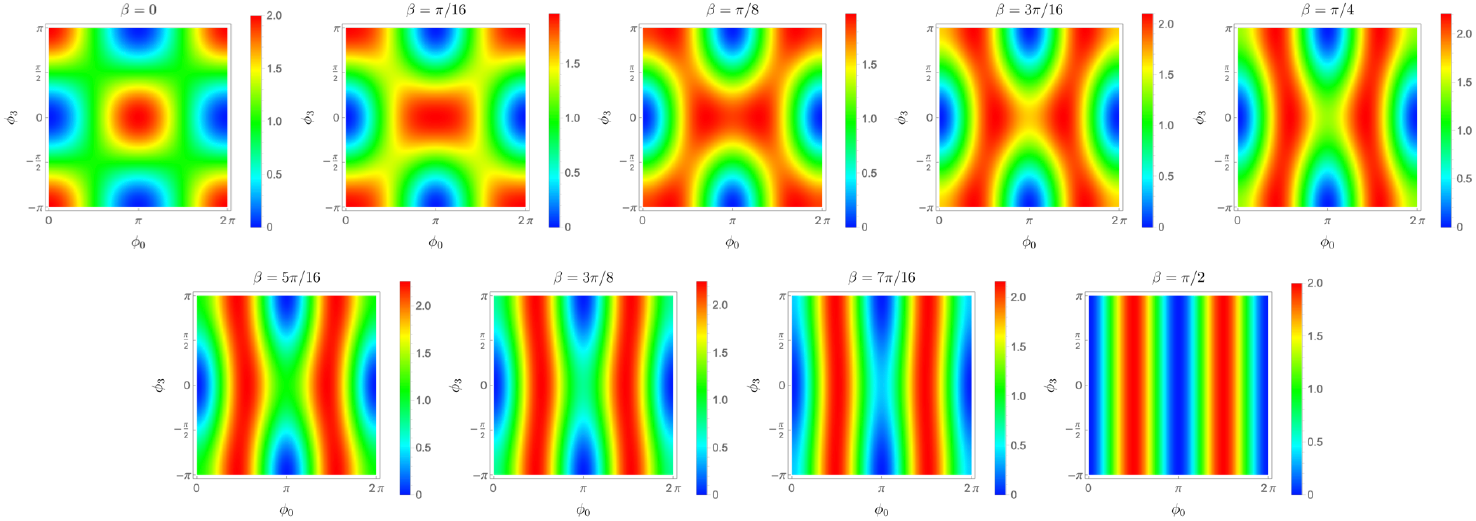}
\caption{
The scalar potential $V(\phi_0,\phi_3;\beta$) in the $\phi_0$-$\phi_3$ plane
($\phi_0 \in [0,2\pi]$ and $\phi_3 \in [-\pi,\pi]$)
for $\beta=0$, $\pi/16$, $\cdots$, $\pi/2$.
The colors show the potential hight as indicated by the color bars.}
\label{fig:pot}
\end{center}
\end{figure}

Next, we explain the scalar potential in the chiral Lagrangian studied in the main text.
The dimensionless scalar potential for $\phi_0$ and $\phi_3$ appearing in the Hamiltonian $C^{-1}{\cal H}$ can be
parametrized by $\beta$ defined by $\tan\beta = A/(4mB)$ as
\be
V(\phi_0,\phi_3;\beta) = \sin\beta(1-\cos2\phi_0) + \cos\beta(1-\cos\phi_0\cos\phi_3).
\ee
This is bounded from below as $V \ge 0$.
Note that the case of $\beta =0$ corresponds to the limit in which the anomaly term is ignored, and
the case of $\beta=\pi/2$ corresponds to the chiral limit 
in which the quark mass terms are absent.
Except for the chiral limit, the potential minima are placed at 
$(\phi_0,\phi_3) = (2m \pi, 2n \pi)$
and $((2m+1)\pi, (2n+1)\pi)$ for $m,n \in \mathbb{Z}$ as shown in Fig.~\ref{fig:pot}. 
All of them identically correspond to the unique vacuum in terms of 
the $U(2)$ field $U$: 
$U = {\bf 1}_2$.
The chiral limit $\beta = \pi/2$ for which the potential $V$ does not depend on
$\phi_3$ reflecting the fact that the Lagrangian is invariant under 
the full chiral symmetry 
$U \to V_L^\dag U V_R$. 
This is explicitly broken to the vector symmetry 
$V_L=V_R$ by the mass term for $\beta \neq \pi/2$.

The equations of motion for $\phi_{0,3}$ under the assumption that $\phi_{0,3}$ is static and depends
only on $\zeta$  read
\be
\phi_0'' - \frac{\cos\beta}{2}\left(\sin(\phi_0+\phi_3) + \sin(\phi_0-\phi_3)\right)
- 2\sin\beta\sin2\phi_0 = 0,
\label{eq:eom0}\\
(1-\epsilon)\phi_3'' - \frac{\cos\beta}{2}\left(\sin(\phi_0+\phi_3) - \sin(\phi_0-\phi_3)\right) = 0, 
\label{eq:eom3}
\ee
where the prime denotes differentiation with respect to
$\zeta$.
Note that 
the CVE term does not appear because it is the topological term.
These are the EOMs which we solved to obtain all the background CSs and CSLs configurations
in the main text.

Note that $\phi_3=0$ solves Eq.~(\ref{eq:eom3}), and Eq.~(\ref{eq:eom0}) reduces to
\be
\phi_0'' - \cos\beta\sin\phi_0 - 2\sin\beta\sin2\phi_0 = 0.
\ee
The second and third terms have different periodicities $2\pi$ and $\pi$ with respect to $\phi_0$, and
the above equation is the so-called double sine-Gordon equations in the literature \cite{PhysRevB.27.474,Ross:2020orc}.
However, solutions ($\eta$-CS and $\eta$-CSL) with $\phi_3=0$ are not necessarily stable
because the pions can be tachyonic for some parameter regions, 
as explained in the main text.

\begin{figure}[t]
\begin{center}
\includegraphics[width=\textwidth]{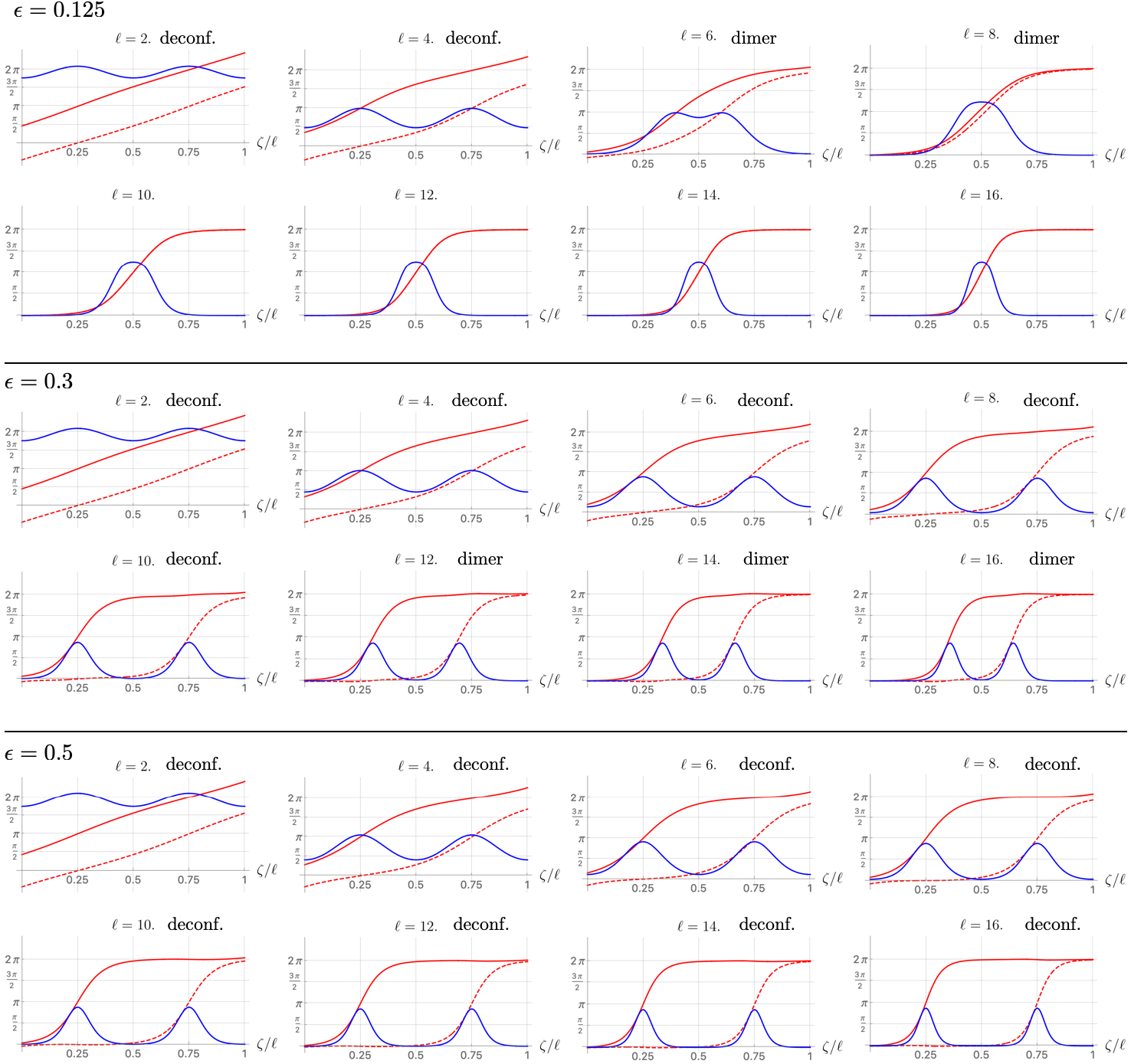}
\caption{The numerical solutions of CSLs with the fixed $\beta=\pi/16$. The solutions in the top/middle/bottom
row are for $\epsilon=$0.125/0.3/0.5, respectively. For each $\epsilon$ we show the CSL with the 
lattice space $\ell = 2,4,6,8,10,12,14,16$. The horizontal axes is normalized coordinate $\zeta/\ell$,
and the red-solid and red-dashed curves show $\phi_+$ and $\phi_-$, respectively. The blue-solid curve
correspond to the energy density. We put ``deconf." if the CSL is in the deconfinement phase,
and ``dimer'' if it is in the dimer phase.}
\label{fig:sols_beta_pi_over_16}
\end{center}
\end{figure}
In Fig.~\ref{fig:sols_beta_pi_over_16} we show three typical series of CSLs with $\beta$ fixed as $\beta = \pi/16$, 
whereas $\epsilon$ is varied as
$\epsilon = 0.125$, $0.3$, and $0.5$. 
As we explain in the main text, $\beta$ and $\epsilon (> 0)$ yield
attractive and repulsive forces 
between u-CS and d-CS, respectively. The repulsive force for $\epsilon=0.125$ is 
relatively weak compared with the attractive force at $\beta=\pi/16$, so that the CSL with 
sufficiently large period $\ell$ is an $\eta$-CSL in the confining phase. As the period $\ell$ decreases, 
the CSL enters in the dimer phase, and finally goes into the deconfing phase.
The repulsive force of $\epsilon = 0.3$ is stronger than that of $\epsilon = 0.125$, so that
the CSL at the large lattice size $\ell$ is not the $\eta$-CSL but the dimer NA-CSL. 
When $\epsilon$ is further large, $\epsilon = 0.5$, the repulsive force always dominates and
the CSL is always in the deconfining phase.
These solutions are used for obtaining Fig.~3 in the main text.

\section{Derivation of Linearized EOMs around the Backgrounds}
\label{app:b}

Here we derive linearized EOMs for linear perturbations 
around background solutions.
Let $\bar\phi_0$ and $\bar\phi_3$ be a solution to Eqs.~(\ref{eq:eom0}) and (\ref{eq:eom3}) with
$\bar\phi_1 = \pi_1/f_\pi = 0$ and $\bar\phi_2 = \pi_2/f_\pi = 0$.

\if
Consider small fluctuations  around the background solution as
\be
\phi_{0,3} = \bar \phi_{0,3}(\zeta) + \delta \phi_{0,3},\quad
\phi_{1,2} = \delta \phi_{1,2}.
\ee

First, the 
linearized EOM for the zeroth and third components are given by
\be
\left[
\p^2
+
\left(
\begin{array}{cc}
4 \sin\beta \cos 2\bar\phi_0 + \cos\beta \cos\bar\phi_0\cos\bar\phi_3 &
- \cos\beta \sin\bar\phi_0\sin\bar\phi_3 \\
\frac{-\cos\beta\sin\bar\phi_0\sin\bar\phi_3}{1-\epsilon} &
\frac{\cos\beta\cos\bar\phi_0\cos\bar\phi_3}{1-\epsilon}
\end{array}
\right)
\right]
\left(
\begin{array}{c}
\delta\phi_0\\
\delta\phi_3
\end{array}
\right)
= 0. \label{eq:linearEOM0}
\ee
It is sometimes useful to change the basis by
\be
\bar\phi_\pm = \bar\phi_0 \pm \bar\phi_3,\quad \delta\phi_\pm = \delta\phi_0 \pm \delta\phi_3.
\ee
Then, Eq.~(\ref{eq:linearEOM0}) can be rewritten as
\be
\left[
\p^2 + 
\left(
\begin{array}{cc}
2\sin\beta\cos2\bar\phi_0 + \frac{2-\epsilon}{2(1-\epsilon)}\cos\beta\cos\bar\phi_+ &
2\sin\beta\cos2\bar\phi_0 - \frac{\epsilon}{2(1-\epsilon)}\cos\beta\cos\bar\phi_- \\
2\sin\beta\cos2\bar\phi_0 - \frac{\epsilon}{2(1-\epsilon)}\cos\beta\cos\bar\phi_+ &
2\sin\beta\cos2\bar\phi_0 + \frac{2-\epsilon}{2(1-\epsilon)}\cos\beta\cos\bar\phi_-
\end{array}
\right)
\right]
\left(
\begin{array}{c}
\delta\phi_+\\
\delta\phi_-
\end{array}
\right) = 0.
\ee

Second,
to write down the linearized EOMs for $\delta\phi_{1,2}$, it is useful to change the variables
as $\delta\pi_{\pm} = \delta\phi_1 \pm i \delta \phi_2$. The resulting equations 
are given by
\be
\left[(1-\epsilon)\left(\p^2 +2i\bar\phi_3'\p_\zeta -i(\bar\phi_3')^2\right) + \cos\beta\cos\bar\phi_0\cos\bar\phi_3\right]\delta\pi_\pm = 0.
\ee
The kinetic term can be canonicalized by
\be 
\delta \tilde\pi_\pm = e^{\mp i \bar\phi_3}\delta\pi_\pm,
\ee
and we get
\be
\left[\p^2 - (\bar\phi_3')^2 + \frac{\cos\beta}{1-\epsilon}\cos\bar\phi_0\cos\bar\phi_3\right]
\delta\tilde\pi_\pm = 0.
\ee
\fi

Consider the derivation of linearized equations of motion for small fluctuations:
\be
\phi_{0, 3} = \bar \phi_{0, 3}(\zeta) + \delta \phi_{0, 3},\quad
\phi_{1, 2} = \delta \phi_{1, 2}.
\ee
For the zeroth and third components, we find
\be
\left[
\p^2
+
\left(
\begin{array}{cc}
4 \sin\beta \cos 2\bar\phi_0 + \cos\beta \cos\bar\phi_0\cos\bar\phi_3 &
- \cos\beta \sin\bar\phi_0\sin\bar\phi_3 \\
\frac{-\cos\beta\sin\bar\phi_0\sin\bar\phi_3}{1-\epsilon} &
\frac{\cos\beta\cos\bar\phi_0\cos\bar\phi_3}{1-\epsilon}
\end{array}
\right)
\right]
\left(
\begin{array}{c}
\delta\phi_0\\
\delta\phi_3
\end{array}
\right)
= 0.
\ee
In some cases, it is useful to change the basis by
\be
\bar\phi_\pm = \bar\phi_0 \pm \bar\phi_3,\quad \delta\phi_\pm = \delta\phi_0 \pm \delta\phi_3.
\ee
In terms of the new basis, the linearized equations are expressed as
\be
\left[
\p^2 + 
\left(
\begin{array}{cc}
2\sin\beta\cos2\bar\phi_0 + \frac{2-\epsilon}{2(1-\epsilon)}\cos\beta\cos\bar\phi_+ &
2\sin\beta\cos2\bar\phi_0 - \frac{\epsilon}{2(1-\epsilon)}\cos\beta\cos\bar\phi_- \\
2\sin\beta\cos2\bar\phi_0 - \frac{\epsilon}{2(1-\epsilon)}\cos\beta\cos\bar\phi_+ &
2\sin\beta\cos2\bar\phi_0 + \frac{2-\epsilon}{2(1-\epsilon)}\cos\beta\cos\bar\phi_-
\end{array}
\right)
\right]
\left(
\begin{array}{c}
\delta\phi_+\\
\delta\phi_-
\end{array}
\right) = 0.
\ee

To obtain the linearized equations for $\delta\phi_{1, 2}$, it is useful to modify the variables
as $\delta\pi_{\pm} = \delta\phi_1 \pm i \delta \phi_2$. The resulting equations 
are decoupled as
\be
\left[(1-\epsilon)\left(\p^2 +2i\bar\phi_3'\p_\zeta -i(\bar\phi_3')^2\right) + \cos\beta\cos\bar\phi_0\cos\bar\phi_3\right]\delta\pi_\pm = 0.
\ee
The kinetic term can be canonicalized by
\be 
\delta \tilde\pi_\pm = e^{\mp i \bar\phi_3}\delta\pi_\pm,
\ee
and we obtain
\be
\left[\p^2 - (\bar\phi_3')^2 + \frac{\cos\beta}{1 - \epsilon}\cos\bar\phi_0\cos\bar\phi_3\right]
\delta\tilde\pi_\pm = 0.
\ee
Thus, $\delta\phi_{0,3}$ ($\delta\phi_\pm$) are mixed to each other, while
$\delta\tilde\pi_\pm$ are isolated.

\subsubsection{$\eta$-CSL}

When the background solution is the $\eta$-CSL, $\bar\phi_3 = 0$, 
the above linearized EOMs reduce to
\be
\left[
\p^2
+
\left(
\begin{array}{cc}
4 \sin\beta \cos 2\bar\phi_0 + \cos\beta \cos\bar\phi_0 &
0 \\
0 &
\frac{\cos\beta\cos\bar\phi_0}{1-\epsilon}
\end{array}
\right)
\right]
\left(
\begin{array}{c}
\delta\phi_0\\
\delta\phi_3
\end{array}
\right)
= 0,
\label{eq:linearEOM}\\
\left(\p^2 + \frac{\cos\beta\cos\bar\phi_0}{1-\epsilon}\right)
\delta\pi_\pm = 0.
\ee
Note that $\delta\phi_3$ and $\delta\pi_\pm = \delta\tilde\pi_\pm\big|_{\bar\phi_3=0}$ 
satisfy the identical equations, which is expected from the fact that the $SU(2)_{\rm V}$ symmetry is kept by the $\eta$-CSL.  
Note that $\delta \phi_0$ is decoupled from the others, and its linearized EOM is
identical to that of the double sine-Gordon equation. Therefore, no tachyonic instabilities
arise in the $\delta\phi_0$ sector. The lowest eigenstate is the translational NG mode whose 
eigenvalue is exactly zero.

On the other hand, the stability in the pion sector can be 
analyzed by
the Schr\"odinger-like equation
\be
\left(- \frac{d^2}{d\zeta^2} + \frac{\cos\beta\cos\bar\phi_0}{1-\epsilon}\right)\psi_n = m_n^2 \psi_n.
\ee
The linear stability of the $\eta$-CSL can be
clarified by obtaining mass square eigenvalues: if the lowest eigenvalue is zero or positive (negative), 
the $\eta$-CSL is (un)stable against local fluctuations. 
We thus obtain a part of 
Fig.~2 (a) and (b) and Fig.~3 (c) and (d) in the main text.

\subsubsection{NA-CSL at $\beta = \epsilon = 0$}
We here study 
the stability of NA-CSLs for 
the special case $\beta = \epsilon = 0$ , where the up-CSL and down-CSL
do not interact.
This can be easily seen by rewriting the Hamiltonian in terms of $\phi_{\pm} = \phi_0 \pm \phi_3$.
It is merely a sum of two sine-Gordon Hamiltonians: ${\cal H} = {\cal H}_+ + {\cal H}_-$ with
\be
C^{-1} {\cal H}_\pm = 
\frac{1}{2}\left\{\frac{1}{2}\left(\frac{d\phi_\pm}{d\zeta}\right)^2 
+ 1 -  \cos\phi_\pm
- S\frac{d\phi_\pm}{d\zeta}\right\},
\label{eq:H_pm}
\ee
and the Schr\"odinger-like equations are
\be
\left(
-\frac{d^2}{d\zeta^2} + 
\cos\bar\phi_\pm
\right)
\delta\phi_{\pm,n} = m_{\pm,n}^2 \delta\phi_{\pm,n},
\label{eq:LEOM03}\\
\left(
-\frac{d^2}{d\zeta^2}- (\bar\phi_3')^2 + \cos\bar\phi_0\cos\bar\phi_3\right)
\delta\tilde\pi_{\pm,n} = \tilde m_{\pm,n}^2 \delta\tilde\pi_{\pm,n}.
\label{eq:LEOM12}
\ee

Let us consider a background CSL with the u- and d-CSLs with the separation $d$ as
\be
\bar\phi_+ = 2\,{\rm am}\left(\frac{\zeta-d/2}{k},k\right) + \pi,\quad
\bar\phi_- = 2\,{\rm am}\left(\frac{\zeta+d/2}{k},k\right) + \pi,
\label{eq:NACSL_back}
\ee
where ${\rm am}(z,k)$ denotes the Jacobi's amplitude function and $k$ is the elliptic modulus taking value within $0 \le k \le 1$.
The $\delta\phi_-$-sector is a just copy of $\delta\phi_+$-sector, and they are identical to 
the linearized EOM of the well-known sine-Gordon soliton. Therefore, we immidiately find the lowest
eigenstate is the massless as 
\be
(m_{\pm,0})^2 = 0,\quad \delta\phi_{\pm,0}={\rm dn} \left(\frac{\zeta\mp d/2}{k},k\right). 
\label{eq:NACSL_phi_pm}
\ee
These two are the independent phonons propagating on the u- and d-CSLs. 
The overall translation is the usual translational zero mode,
while the separation $d$ between them corresponds to 
the so-called quasi-NG mode 
 \footnote{
 Quasi-NG modes are gapless modes which are 
 not related to symmetry breaking of the whole system 
 \cite{Nitta:1998qp,Nitta:2014jta}. 
 Quasi-NG modes associated to the presence of parallel domain walls were found in Ref.~\cite{Eto:2008dm}. In this case, the exchange between NG and quasi-NG modes 
 with keeping the total number occurs 
 \cite{Nitta:1998qp} 
 when two solitons coincide.
 In the NA-CSL for $\beta = \epsilon = 0$, 
 there are three NG modes (translation and isospinons) 
 and one quasi-NG mode 
  when the u-CSL and d-CSL are separated,
 while there are 
 one NG mode (translation) 
 and three quasi-NG modes when they coincide.
 }
which eventually exists  for 
the non-interactive case 
 $\beta = \epsilon = 0$ and 
 becomes gapped in the general case.

On the other hand, 
the Schr\"odinger potential for $\delta\tilde\pi_\pm$ is more complicated.
It can be explicitly written as
\be
- (\bar\phi_3')^2 + \cos\bar\phi_0\cos\bar\phi_3
=  - \frac{1}{k^2}\left(
{\rm dn}_+ - {\rm dn}_-\right)^2
- {\rm cn}_+^2 {\rm cn}_-^2
+ {\rm sn}_+^2 {\rm sn}_-^2,
\ee
with 
${\rm dn}_\pm = {\rm dn}\left(\frac{\zeta \pm d/2}{k},k\right)$,
and similar to ${\rm cn}$ and ${\rm sn}$, 
where ${\rm dn}$, 
 ${\rm cn}$ and ${\rm sn}$ are the Jacobi's elliptic functions;
the delta amplitude,
the elliptic cosine, and the elliptic sine, respectively.
We find that the lowest eigenvalues are zero and
the corresponding mode functions are given by
\be
(\tilde m_{\pm,0})^2 = 0,\quad
\delta\tilde \pi_{\pm,0} = {\rm dn} \left(\frac{\zeta - d/2}{k},k\right)
+ {\rm dn} \left(\frac{\zeta + d/2}{k},k\right).
\label{eq:NACSL_pi_pm}
\ee
These correspond to the gapless NG modes, 
isospinons explained in the main text, 
associated with the symmetry breaking $SU(2)_{\rm V} \to U(1)$ in the presence of the NA-CSL. 
These isospinons are type-A NG modes having linear dispersion relations  
in the classification of NG modes 
\cite{Watanabe:2012hr,Hidaka:2012ym,Takahashi:2014vua}
since they are antiffero. 
While NA NG modes propagatng on a NA vortex lattice are known \cite{Kobayashi:2013axa}, those on 
a soliton (domain wall) lattice found in this Letter are new.

\end{document}